\documentclass[a4paper,12pt]{article}            
\usepackage{jheppub}
\pdfoutput=1

\usepackage[utf8]{inputenc}
\usepackage{amsmath}
\usepackage{mathrsfs}  
\usepackage{tensor}
\usepackage{bbm}
\usepackage{multirow}
\usepackage{enumitem}
\usepackage[normalem]{ulem}
\allowdisplaybreaks
\usepackage{comment}

\title{Non-dissipative electrically driven fluids}
\author[1,2]{Andrea Amoretti,}
\author[3]{Daniel K. Brattan,} 
\author[1,2]{Luca Martinoia,}
\author[1,2]{Ioannis Matthaiakakis.}

\emailAdd{andrea.amoretti@ge.infn.it}
\emailAdd{danny.brattan@gmail.com}
\emailAdd{luca.martinoia@ge.infn.it}
\emailAdd{Ioannis.Matthaiakakis@edu.unige.it}

\affiliation[1]{Dipartimento di Fisica, Universit\`a di Genova,
via Dodecaneso 33, I-16146, Genova, Italy}
\affiliation[2]{I.N.F.N. - Sezione di Genova, via Dodecaneso 33, I-16146, Genova, Italy}
\affiliation[3]{CPHT, \'{E}cole Polytechnique, 91128 Palaiseau cedex, France}

\begin{abstract}
{\ Existing hydrodynamic models of charged fluids consider any external electric field acting on the fluid as either first order in the hydrodynamic derivative expansion and completely arbitrary or zeroth order but constrained by the fluid's chemical potential. This is in tension with experiments on charged fluids, where the electric field is both zeroth order and completely arbitrary. In this work, we take the first step at resolving this conundrum by introducing a new class of hydrodynamic stationary states, including an arbitrary zeroth order electric field, upon which hydrodynamics can be built. We achieve this by first writing down the hydrostatic constitutive relations for a boost-agnostic charged fluid up to first order in derivatives. Then we introduce suitable energy and momentum relaxation terms to balance the influence of the electric field on the fluid. This analysis leads to a new hydrostatic constraint on the spatial fluid velocity, which can be used to define our class of states. This constraint generalizes to the realm of hydrodynamics a similar constraint on the velocity found in the Drude model of electronic transport. Our class of states exhibits non-trivial thermo-electric transport even at ideal order, since it hosts non-zero DC electric and heat currents. We derive the explicit form of the corresponding conductivities and show they depend non-linearly on the electric field.}
\end{abstract}

\begin{document}

\maketitle

\section{Introduction}\label{sec:Intro}

{\noindent The prototypical example of transport experiments in fluids most, if not all, of us are accustomed to is that of charge transport across a piece of wire. Stripped to its essential form, this experiment involves a charged fluid forced to flow by an external electric field $\vec{\mathbbm{E}}$ where the electric field is always assumed to be external and linearised in amplitude.}

{\ In the absence of non-conservation of momentum, stability of the system requires that we constrain $\vec{\mathbbm{E}}$ in terms of the derivative of the chemical potential $\mu$. In this case the background electric field effectively disappears from the dynamics, cancelled by the chemical potential gradient in hydrostatic equilibrium \cite{Kovtun:2016lfw}, and therefore the fluid velocity is independent of the applied electric field. This is not the case in real devices which are open systems and, when attached to a voltage bias so that a current flows, achieve stationary states by exchanging heat with their surroundings. In this work, we resolve this paradox for the case of boost-agnostic fluids \cite{deBoer:2017ing,deBoer:2017abi,Novak:2019wqg,Armas:2019gnb,deBoer:2020xlc,Armas:2020mpr} and introduce stationary states where an \textit{arbitrary} external field and fluid polarization can co-exist. In particular, we find a new class of hydrodynamic stationary states for which the electric field is not completely opposed by a chemical potential gradient as well as order zero in the hydrodynamic derivative expansion. Importantly, these states exhibit charge and heat transport in the ground state, indicating they provide the correct description for standard DC measurements. Furthermore, these stationary states provide the basis upon which we can include dissipative corrections to the hydrodynamic constitutive relations, upon breaking stationarity. } 

{\ To achieve the goal of this paper, we must relax some familiar concepts from recent studies of hydrodynamics. For example, turning on an order zero electric field pushes the fluid out of equilibrium, by continuously pumping energy and momentum into it. To avoid a ``blow-up", while keeping the electric field an externally imposed parameter that is not completely balanced by the chemical potential gradient, we must introduce order-zero sinks for the fluid energy and momentum. In addition, the order zero sinks necessarily break boost invariance. Hence, our use of the boost-agnostic formalism \cite{deBoer:2017ing,deBoer:2017abi,Novak:2019wqg,Armas:2019gnb,deBoer:2020xlc,Armas:2020mpr} is not a choice, but a requirement for a consistent hydrodynamic theory. As a result, the fluid's spatial velocity $\vec{v}$---which can be set to zero in boost invariant hydrodynamics---is now a parameter of thermodynamic equilibrium. It follows that our theory of hydrodynamics describes fluids with charge and momentum flow at zeroth order in the derivative expansion, contrary to expectations from the boost-invariant formalism \cite{Kovtun:2012rj,Kovtun:2016lfw}. With the addition of the energy-momentum sinks, we can conceptualize our approach as the hydrodynamic analogue of the Drude model of electron transport \cite{AshcroftMermin}.}

{\ We wish to remain completely agnostic of the microscopic origins of our energy or momentum sinks. Hence, our hydrodynamic description is applicable to as broad a range of systems as possible with a single exception; we assume that the corresponding susceptibilities given by varying with respect to the ``hidden'' degrees of freedom represented by the relaxation terms are negligible. Consequently, these degrees of freedom are frozen out and do not partake in the thermodynamics (nor hydrodynamics) of the system except through relaxation of the hydrodynamic charges and currents. Put differently, the degrees of freedom represented by the relaxation rates act somewhat like a bath for our fluid. It is important to note however that the relaxation rates can depend on the thermodynamic variables $T$, $\mu$, $\vec{v}$ and $\vec{\mathbbm{E}}$; it is only the hidden degrees of freedom (for example a translation breaking scalar) that our hydrodynamics must not depend upon. This is similar to the situation in \cite{Landry:2020tbh} in the context of charge relaxations, but opposite to the approach than \cite{Becattini:2208}, in which the environment is taken to completely dominate the thermodynamics of the system. In addition \cite{Visuri:2209} discusses the steady state of an open system that loses particles to the environment, albeit in a non-hydrodynamic approach.

{\ Recall that within the Drude model, a steady state is reached only when the flow velocity is constrained in terms of the rest of the system's parameters. This feature of the Drude model survives the hydrodynamic generalization, where we find that a stationary state can be reached only when $\vec{v}$ is constrained as 
	\begin{eqnarray}
		\label{Eq:VelocityRelation}
		v^{i} &=& \Omega_{\mathbbm{E}}\left(T,\mu,\vec{v}^2,\vec{\mathbbm{E}}^2,\vec{v} \cdot \vec{\mathbbm{E}}\right) \mathbbm{E}^{i} - \Omega_{\mu}\left(T,\mu,\vec{v}^2,\vec{\mathbbm{E}}^2,\vec{v} \cdot \vec{\mathbbm{E}}\right) \partial^{i} \mu \; , 
	\end{eqnarray}
where $T$ is the fluid's temperature and $\Omega_{\mathbb{E},\mu}$ are functions of the thermodynamical variables fixed in later sections. Note that we included a chemical potential term on the right hand side of \eqref{Eq:VelocityRelation}, because of the ability of $\partial^i\mu$ to act as an effective electric field. The derivation of Eq.\eqref{Eq:VelocityRelation} follows the natural and normative way familiar from Lagrangian mechanics; namely, we derive the conservative part of the system from a variational principle and then add non-conservative forces by hand and enforce consistency with the equations of motion and second law of thermodynamics. We see that the result of this analysis is a velocity whose value can be chosen at will by fixing the rest of the thermodynamic parameters. In this way, our formalism mimics the boost-invariant formalism and explains why boost-invariant hydrodynamics has been successfully used to describe condensed matter experiments (see e.g. \cite{Gooth2018,PhysRevB.100.245305,Sulpizio2019,Amoretti:2019buu,Ku2020,Kumar2022}).

{\ One might object that by having such relaxation rates at order zero, our hydrodynamic modes will become strongly decaying. Before addressing this issue, let us clarify the meaning of strong and order zero. There are two senses of small which may be encountered in hydrodynamics---small in amplitude and small in gradients. A quantity which is small in amplitude need not be small in gradients and vice versa. Consequently, while our decay rates are strong in the derivative sense, they can be small (even perturbatively small) in the amplitude sense. In this way, our relaxation terms follow the traditional route in the hydrodynamics literature. We emphasize this route is essential, for otherwise hydrodynamics could not accurately describe flows in the presence of zeroth order backgrounds fields, e.g. the flow of a fluid in a zeroth order, constant, external magnetic field. For the cyclotron modes of such a theory to be amenable to a hydrodynamic description, i.e. to avoid the formation of Landau levels, the constant magnetic field must be suitably small in amplitude even though it is large in the derivative sense. See also \cite{Delacretaz:2019wzh,Amoretti:2020mkp,Amoretti:2021fch,Amoretti:2021lll,Amoretti:2022acb} for further discussions about external strong magnetic field in boost-invariant hydrodynamics .}

{\ Furthermore, hydrodynamic modes become decaying in many circumstances (external magnetic fields see e.g. \cite{Hartnoll:2007ih}), charge density waves (see e.g. \cite{Amoretti:2019kuf,Armas:2020bmo}), Wigner crystals (see e.g. \cite{Delacretaz:2017zxd,Delacretaz:2019wzh,Armas:2022vpf}), when considering non-zero wavevectors $\vec{k}$ (see e.g. \cite{Kovtun:2012rj} etc.) and yet we expect a hydrodynamic framework to be a good approximation. What matters is not that the mode is strongly decaying, but that the non-hydrodynamic modes are sufficiently deep in the complex plane such that they can be approximately ignored. This can only be stated analytically by knowing the position of the lowest lying non-hydrodynamic pole. More pragmatically, we can simply check whether the hydrodynamic and measured AC conductivity agree with each other. Regardless, if there is any hope for hydrodynamics to explain standard DC transport measurements as discussed above, then there must be a regime where such non-hydrodynamic modes are irrelevant. The analysis presented in this work presumes such systems exist.}

{\ As a last comment, we note that the fluids we consider are stationary since there is no entropy production. This can be demonstrated explicitly to be a consequence of the (non-)conservation equations for energy, momentum and charge. When we discuss positivity of the entropy current we will also see that this condition follows from requiring that the ideal fluid does not produce entropy.}

{\ We proceed to the main part of this paper in the following manner: we first derive the boost-agnostic hydrostatic constitutive relations, using the generating functional technique, in the absence of any explicit energy or momentum relaxation in section \ref{sec:IdealHydro}. In the same section, we relax conservation of our hydrodynamic charges and impose \eqref{Eq:VelocityRelation} on the system. In doing so we determine $\Omega_{\mathbb{E},\mu}$ in \eqref{Eq:VelocityRelation} in terms of hydrodynamic variables and generalize the Drude result to hydrodynamic systems. Following that, we consider relaxation terms at order one in derivatives in section \ref{sec:1stOrderHydro} showing how our results must be extended and modified before obtaining the DC thermoelectric conductivities in section \ref{sec:Conductivities}. We conclude with a discussion of our results and their applications in section \ref{sec:Discussion}.}

\section{Stationarity at the ideal level}\label{sec:IdealHydro}

{\noindent In the present section, we use the generating functional formalism \cite{PhysRevLett.109.101601, Banerjee2012,Bhattacharya2013} to write down the zeroth order constitutive relations for the stress-energy-momentum tensor and $U(1)$ charge current of our boost-agnostic charged fluids in hydrostatic equilibrium and in the presence of an external electric field $\vec{\mathbb{E}}$. To use the hydrostatic generating functional we need to geometrise the thermodynamics of the fluid. Since boosts are not necessarily part of our fluid's symmetries, we must formulate the problem in a boost-agnostic manner. Consequently, we turn to Aristotelian geometry \cite{deBoer:2017ing,deBoer:2017abi,Novak:2019wqg,Armas:2019gnb,deBoer:2020xlc,Armas:2020mpr}, which we review in the next subsection. We proceed in subsection \ref{subsec:GeomThermo} with defining the thermodynamic fluid parameters and the conditions they need to satisfy for the fluid to be in hydrostatic equilibrium. With the above tools in hand, we construct the generating functional and derive the fluid's constitutive relations in \ref{subsec:HSConRelations}. We finally relax the hydrostatic constraints and write down the hydrostatic constitutive relations for generic energy-momentum relaxation in subsection \ref{subsec:0thOrderRelaxation}.}

\subsection{Aristotelian geometry}\label{subsec:AristoGeom}

{\noindent Aristotelian geometry consists of a manifold equipped with a clock form $\tau_\mu$ and a spatial metric $h_{\mu\nu}$, the latter of which has the signature $(0,1,\dots,1)$. Besides lacking boost-invariance, another reason for considering this geometry is that it naturally allows us to separate out the spatial part of the fluid velocity by defining a laboratory frame aligned with $\tau_\mu$.  Thus we can define fluid frame invariant notions of energy and momentum such as $\tau_{\mu} T\indices{^\mu_\nu}$. This is distinct from the usual relativistic case where the dual to the time translation vector is the fluid velocity one-form $u_{\mu}$ and there is no obvious way to isolate the spatial part of this velocity field from its temporal counterpart except by introducing a field exactly like $\tau_{\mu}$.\footnote{This is usually achieved by working in Cartesian coordinate system, with $\tau_{t}= 1$, and identifying $u^{\mu} = (1,v^{i}) / \sqrt{1-\vec{v}^2}$.} Moreover, Lorentzian, Galilean, Lifshitz and other geometries are a special limit of these Aristotelian setups.}

{\ It will simplify some expressions if we assume that we can decompose the spatial metric $h_{\mu \nu}$ with respect to vielbeins i.e.
\begin{equation}
    h_{\mu\nu}=\delta_{ab}e^a_\mu e^b_\nu ~~ ,~~ a,b= 1,2,...d
\end{equation}
with $d$ the number of spatial dimensions. Subsequently, we assume that the square matrix $(\tau_\mu,e^a_\mu)$ is invertible and thus obtain $(-\nu^\mu, e^\mu_a)$ obeying the following conditions
\begin{equation}
    \nu^\mu\tau_\mu=-1 \; , \qquad \nu^\mu e^a_\mu=0 \; , \qquad e^\mu_a\tau_\mu=0 \; , \qquad e^\mu_ae^b_\mu=\delta^b_a \; , 
\end{equation}
and the completeness relation
\begin{equation}
   \label{Eq:Vielbeincompleteness}
    -\nu^\mu\tau_\nu+e^\mu_a e^a_\nu=\delta^\mu_\nu \; . 
\end{equation}
We can also define
\begin{equation}
    h^{\mu\nu}=\delta^{ab}e^\mu_ae^\nu_b
\end{equation}
which is \underline{not} the inverse of $h_{\mu\nu}$, but satisfies
\begin{equation}
    \label{Eq:DecompositionofIdentity}
    h_{\mu\rho}h^{\rho\nu}=\delta^\nu_\mu+\nu^\nu\tau_\mu \; .
\end{equation}
As we see from right-hand side of Eq. \eqref{Eq:DecompositionofIdentity}, $h_{\mu\rho}h^{\rho\nu}$ is a projector normal to $\tau_\mu$. Thus we can think of $h^{\mu\nu}$ as the inverse of $h_{\mu\nu}$ \textit{only} on the spatial hypersurfaces defined by the clock-form $\tau_\mu$.

We introduce a metric compatible covariant derivative into our spacetime, constrained via the following properties:
	\begin{eqnarray}
		\nabla_{\mu} \tau_{\nu} = 0 \; , \qquad \nabla_{\mu} h^{\nu \rho} = 0 \; .
	\end{eqnarray}
Connections compatible with these conditions have the form
	\begin{eqnarray}
		\label{Eq:Genericconnection}
		\Gamma^{\lambda}_{\mu \nu} &=& - \nu^{\lambda} \partial_{\mu} \tau_{\nu} + \frac{1}{2} h^{\lambda \kappa} \left( \partial_{\mu} h_{\nu \kappa} + \partial_{\nu} h_{\mu \kappa} - \partial_{\kappa} h_{\mu \nu} \right) + \frac{1}{2} h^{\lambda \sigma} Y_{\sigma \mu \nu} \; ,
	\end{eqnarray}
where $Y_{\sigma \mu \nu}$ is a otherwise arbitrary tensor \cite{Hartong:2015zia} satisfying
	\begin{eqnarray}
		\left( h^{\rho \sigma} h^{\lambda \nu} - h^{\lambda \sigma} h^{\rho \nu} \right) Y_{\sigma \mu \nu} =  0 \; .
	\end{eqnarray}
One can check explicitly that \eqref{Eq:Genericconnection} transforms as a connection if $Y_{\sigma \mu \nu}$ transforms like a tensor. Because $Y_{\sigma \mu \nu}$ is arbitrary, so is the connection. This should be compared to the usual pseudo-Riemannian case, where the metric uniquely determines a preferred connection (the Levi-Civita connection). Consequently, the physics in a curved Aristotlean spacetime can depend on the arbitrary choice of connection.}	

{\ To fix our choice of a connection, we also require metric compatibility of the dual vector $\nu^\mu$ and spatial inverse $h^{\mu\nu}$, i.e.
	\begin{eqnarray}
		\label{Eq:Metriccompatible}
		\nabla_{\mu} \nu^{\nu} = 0 \; , \qquad \nabla_{\mu} h_{\nu \rho} = 0 \; .
	\end{eqnarray}
This choice makes identifying independent scalars in our generating functional significantly easier. Following \cite{deBoer:2020xlc}, a suitable ansatz  for the affine connection has the form
	\begin{eqnarray}
		\Gamma^{\lambda}_{\mu \nu} &=& - \nu^{\lambda} \partial_{\mu} \tau_{\nu} + \frac{1}{2} h^{\lambda \kappa} \left( \partial_{\mu} h_{\nu \kappa} + \partial_{\nu} h_{\mu \kappa} - \partial_{\kappa} h_{\mu \nu} \right) - h^{\lambda \kappa} \tau_{\nu} K_{\mu \kappa} + C_{\mu \nu}^{\lambda} 
	\end{eqnarray}
where $K_{\mu \nu}$ is the extrinsic curvature defined by
	\begin{eqnarray}
		K_{\mu \nu} = - \frac{1}{2} \mathcal{L}_{\nu} h_{\mu \nu}	\;  
	\end{eqnarray}
\noindent and
	\begin{eqnarray}
		C_{\mu \nu}^{\lambda} \tau_{\lambda} = 0 \; , \qquad C_{\mu \nu}^{\lambda} h_{\lambda \rho} + C_{\mu \rho}^{\lambda} h_{\nu \lambda} = 0 \; . 
	\end{eqnarray}
A particularly simple solution for $C^\lambda_{\mu\nu}$ is taking $C_{\mu \nu}^{\lambda} \equiv 0$ i.e.
	\begin{eqnarray}
		\label{Eq:Ourconnection}
		\Gamma^{\lambda}_{\mu \nu} &=& - \nu^{\lambda} \partial_{\mu} \tau_{\nu} + \frac{1}{2} h^{\lambda \kappa} \left( \partial_{\mu} h_{\nu \kappa} + \partial_{\nu} h_{\mu \kappa} - \partial_{\kappa} h_{\mu \nu} \right) - h^{\lambda \kappa} \tau_{\nu} K_{\mu \kappa} \; . 
	\end{eqnarray}
	
To conclude our short overview of Aristotelian geometry, we note that in our current work we are interested in ``flat spacetimes'' i.e. ones where the elementary tensor structures reduce to
\begin{equation}
    \tau_\mu=\delta^0_\mu \; , \qquad h_{\mu\nu}=\delta^i_\mu\delta^j_\nu \delta_{ij} \; , \qquad \nu^\mu=-\delta^\mu_0 \; , \qquad h^{\mu\nu}=\delta^\mu_i\delta^\nu_j \delta^{ij} \; . 
\end{equation}
In particular, our choice of tensor structure physically means that laboratory time is aligned with $\tau_{\mu}$ and that $h^{\mu\nu}$ can be used as a spatial Euclidean metric. Essentially, in the flat space limit, we generally reduce to a Cartesian coordinate system where $\nabla_{\mu} = \partial_{\mu}$ so that
	\begin{eqnarray}
		\partial_{\mu} \tau_{\nu} = 0 \; , \qquad \partial_{\mu} h_{\nu \rho} = 0 \; . 
	\end{eqnarray}
We shall call this the flat space cartesian co-ordinates or FSCC limit for short. In the FSCC limit, the precise choice of connection becomes a moot point regarding the final result. However, having the curved space expressions, albeit not the most general such expressions, is necessary for computing the variation of the hydrostatic generating functional. More precisely, notice that our connection of choice \eqref{Eq:Ourconnection} has non-zero torsion, $\Gamma^{\rho}_{[\mu \nu]} \neq 0$, which contributes to our first order generating functional.\footnote{For example, when moving from partial to covariant derivatives, we can generate additional torsion terms such as
	\begin{eqnarray}
		\nabla_{\mu} V^{\mu} &=& \frac{1}{e} \partial_{\mu} \left( e V^{\mu} \right) + \Gamma_{[\mu \nu]}^{\nu} V^{\mu} \; .
\end{eqnarray}
where $e=\text{det}(\tau_\mu,e^a_\mu)$.
}

\subsection{Geometrising thermodynamics}\label{subsec:GeomThermo}

{\noindent We can now express our thermodynamic quantitites in terms of the geometric quantities ($\tau_{\mu}$, $h_{\mu \nu}$, $A_{\mu}$). First, we must define the notion of stationarity. To this end, we introduce the time-direction associated to the fluid's dynamical evolution  in terms of the thermal vector $\beta^\mu$.\footnote{E.g. in terms of Hamiltonian mechanics, where dynamical evolution is given by the Poisson bracket with the fluid Hamiltonian $H$, we have $\lbrace H, \bullet\rbrace = -\beta^\mu\partial_\mu$.} We require $\beta^\mu$ to play the role of a Killing vector in our spacetime. That is the Lie derivative with respect to $\beta^\mu$, $\mathcal{L}_\beta$, vanishes when acting on any geometric object in the theory. When this condition holds, our theory is stationary with respect to $\beta^\mu$.\footnote{Note that due to the lack of boost invariance, stationarity with respect to $\beta^\mu$ does not imply stationarity with respect to any other vector.}  In particular, acting on the sources we have
	\begin{subequations}
		\label{Eq:GeometryConstraints}
		\begin{eqnarray}
			\label{Eq:Killingtau}
    			\mathcal{L}_\beta\tau_\mu &=& 0 \; , \\
			\label{Eq:KillingA}
    			\mathcal{L}_\beta A_\mu &=& 0 \; , \\
			\label{Eq:Killingh}
    			\mathcal{L}_\beta h_{\mu\nu} &=& 0 \; , 
		\end{eqnarray}
	\end{subequations}
as constraints on our geometry in addition to the usual Bianchi identity,
	\begin{eqnarray}
		\label{Eq:BianchiIdentity}
		\partial_{\left[ \mu \right.} F_{\left. \nu \rho \right]} = 0 \; . 
	\end{eqnarray}
}

Up next, we define the temperature $T$ and chemical potential $\mu$ of the fluid as
	\begin{subequations}
	\begin{eqnarray}
		\label{Eq:tempdef}
		T &=& \frac{1}{\tau_{\mu} \beta^{\mu}} \; , \\
		\mu &=& T \left( A_{\mu} \beta^{\mu} + \Lambda_{V} \right) \; ,
	\end{eqnarray}
	\end{subequations} 
respectively, where $\Lambda_{V}$ represents the choice of $U(1)$ gauge. Meanwhile, we take the fluid velocity to be proportional to the Killing vector and normalise it by setting
	\begin{eqnarray}
		\label{Eq:TimeComponentu}
		u^{\mu} \tau_{\mu} = 1 \; , 
	\end{eqnarray}
such that, with \eqref{Eq:tempdef}, we can identify
	\begin{eqnarray}
		u^{\mu} = T \beta^{\mu} \; . 
	\end{eqnarray}
Employing the completeness relation \eqref{Eq:Vielbeincompleteness} and the normalisation condition, \eqref{Eq:TimeComponentu}, we can also decompose the fluid velocity into
	\begin{eqnarray}
		u^{\mu} = - \nu^{\mu} + v^{a} e_{a}^{\mu} \; , \qquad v^{\mu} = u^{\mu} e_{\mu}^{a}
	\end{eqnarray}
with $\vec{v}=v^{a} e_{a}^{\mu} \partial_{\mu}$ the spatial velocity. In the FSCC limit, the velocity reduces to $u^\mu=(1,v^i)$. }

{\  We define the electric field in the laboratory frame as
\begin{equation}
    \label{Eq:Electricfielddef}
     \mathbbm{E}_\mu=-F_{\mu\nu} \nu^\nu \; ,\qquad F_{\mu\nu}= 2 \partial_{[\mu} A_{\nu]} = \mathbbm{E}_\mu\tau_\nu-\mathbbm{E}_\nu\tau_\mu \; , \qquad \mathbbm{E}_{a} = e_{a}^{\mu} \mathbbm{E}_{\mu} \; ,  
\end{equation}
which in flat space with a suitable coordinate choice takes the form $\mathbbm{E}_\mu=(0,\mathbbm{E}_i)$. 

The fact that the electric field can be constant in the reference frame of the laboratory observer is one manner in which boost invariance is broken (see \cite{Amoretti:2022acb} for an explanation in the case of relativistic fluids). Note, that our electric-field definition is different than the one typically encountered in the relativistic fluid literature, where $F_{\mu \nu} u^{\nu} = E_{\mu}$. The electric field $E_{\mu}$ is entirely transverse to the fluid velocity, unlike in our scenario.}

{\ With these definitions for the thermodynamic variables we can reformulate the hydrostatic constraints \eqref{Eq:GeometryConstraints} in terms of the thermodynamic parameters. First, we consider the hydrostatic constraint on the temperature $T$ in Eq.~\eqref{Eq:tempdef}. With the help of \eqref{Eq:Killingtau}, we find 
	\begin{equation}
  \label{Eq:Killing1}
    \frac{\partial_\mu T}{T}-u^\nu\left(\partial_\nu\tau_\mu-\partial_\mu\tau_\nu\right)=0 \; . 
\end{equation}
We can make this expression manifestly covariant by rewriting everything in terms of covariant derivatives as
	\begin{equation}
		 \label{Eq:Torsion1}
		2 u^{\nu} \Gamma_{[\nu \mu]}^{\rho} \tau_{\rho} = \frac{\nabla_{\mu} T}{T} \; . 
	\end{equation}
This result is intuitively correct, given the role of torsion as a deformation of the translation algebra $[\nabla_\mu, \nabla_\nu] = - T^\rho_{\mu\nu}\nabla_\rho$. Essentially, Eq. \eqref{Eq:Torsion1} tells us that a deformation of the time direction, i.e. of temperature, due to the flow can be compensated by a deformation in the translation group algebra, i.e. torsion. Equation \eqref{Eq:Torsion1} also tells us that torsion is not an independent quantity in the hydrostatic limit. We use this observation when working at first order in derivatives in section \ref{sec:1stOrderHydro} to eliminate torsion terms from our list of independent scalars entering the first order generating functional. In the FSCC limit, Eq.~\eqref{Eq:Killing1} simplifies considerably 
\begin{equation}
    \label{Eq:TempConstraints1}
    \partial_{\mu} T = 0 \; . 
 \end{equation}}

{\ Up next, we consider the hydrostatic constraints on the chemical potential $\mu$ \eqref{Eq:tempdef} . Employing the Killing constraint \eqref{Eq:KillingA} on the gauge field and the definition of the electric field \eqref{Eq:Electricfielddef}, we determine that
\begin{equation}
   \label{Eq:KillingAexpand}
    \mathbbm{E}_\mu-T\partial_\mu\frac{\mu}{T}=\mathbbm{E}_\nu u^\nu\tau_\mu \;. 
 \end{equation}
This constraint shows us that the $\partial_{\mu} \mu$ and its derivatives are not independent variables in hydrostatic equilibrium, as they can always be expressed in terms of the electric field $\mathbbm{E}_{\mu}$ and temperature. In FSCC, the set of constraints represented by \eqref{Eq:KillingAexpand} can be written as
	\begin{subequations}
	\label{Eq:ElectricFieldConstraints}
	\begin{eqnarray}
	  \partial_t\mu+v^i\partial_i\mu &=&0 \; , \\
	  \label{Eq:KovtunEcondition}
	  \vec{\mathbbm{E}} \cdot \vec{v} + \partial_t \mu  &=& 0 \; , \\
	  \label{Eq:KovtunEicondition}
	  \vec{\mathbbm{E}} -  \vec{\partial} \mu &=& 0 \; ,
	\end{eqnarray}
	\end{subequations}
where we have employed \eqref{Eq:TempConstraints1}. These are not all independent and, for example, the second constraint is easily recovered from the first and third. Their physical content is thus two-fold, i) the chemical potential is conserved along the flow and ii) the applied electric field $\vec{\mathbb{E}}$ must be balanced by a gradient for the chemical potential $-\vec{\partial}\mu$ in hydrostatic equilibrium. These constraints imply that the chemical potential gradient is actually order zero in derivatives. This is the boost-agnostic version of what already happens in relativistic theories \cite{Kovtun:2016lfw}.}

{\ We now turn to the hydrostatic constraint on the electric field $\mathbb{E}_\mu$ \, i.e. $\mathcal{L}_{\beta}(\mathbbm{E}_{\mu}) =0$. If we had taken the electric field to be order one in derivatives, this constraint would not be relevant to the derivative order we are interested in in this paper. This is because the resulting constraint would then be of order two. As our electric field is zeroth order in derivatives, constancy along $\beta^\mu$ yields
	\begin{eqnarray}
	  0
        &=& u^{\nu} \partial_{\nu} \mathbbm{E}_{\mu} + \mathbbm{E}_{\nu} \partial_{\mu} u^{\nu} - \frac{\mathbbm{E}_{\nu} u^{\nu}}{T} \partial_{\mu} T \; . 
	\end{eqnarray}
Explicitly covariantising this expression we obtain
	\begin{eqnarray}
		\label{Eq:Torsion2}
		2 u^{\nu} \Gamma_{[\nu \mu]}^{\rho} \mathbbm{E}_{\rho}
        &=& u^{\nu} \nabla_{\nu} \mathbbm{E}_{\mu} + \mathbbm{E}_{\nu} \nabla_{\mu} u^{\nu} - \frac{\mathbbm{E}_{\nu} u^{\nu}}{T} \nabla_{\mu} T
	\end{eqnarray}
which again we can use to eliminate torsion terms in our first order action. Moving to flat space we readily find
\begin{eqnarray}
		\label{Eq:Electricfieldconstraints}
		0 &=& \partial_{t} \mathbbm{E}_{i} + v^{j} \partial_{j} \mathbbm{E}_{i} +  \mathbbm{E}_{j} \partial_{i} v^{j}\; , \qquad \mathbbm{E}_{i} \partial_{t} v^{i} = 0 \; . 
\end{eqnarray}
Additionally, the electric field $\mathbbm{E}_{\mu}$ must obey the Bianchi identity $\mathrm{d} F =0$. This has the general form
	\begin{eqnarray}
	  \label{Eq:BianchiIdentityII}
	  0 &=& \partial_{\left[ \mu \right.} \mathbbm{E}_{\left. \nu \right]} \tau_{\rho} + \mathbbm{E}_{\nu} \partial_{\left[ \mu \right.} \tau_{\left. \rho \right]} 
	  	    +  \partial_{\left[ \nu \right.} \mathbbm{E}_{\left. \rho \right]} \tau_{\mu} + \mathbbm{E}_{\mu} \partial_{\left[ \rho \right.} \tau_{\left. \nu \right]} 
	     	    + \partial_{\left[ \rho \right.} \mathbbm{E}_{\left. \mu \right]} \tau_{\nu} + \mathbbm{E}_{\rho} \partial_{\left[ \nu \right.} \tau_{\left. \mu \right]} \; .
	\end{eqnarray}
Covariantising this experssion leads to
	\begin{eqnarray}
	\label{Eq:BianchiIdentityIII}
	   	  \mathbbm{E}_{\sigma} \Gamma^{\sigma}_{[\nu \mu]} \tau_{\rho} + \mathbbm{E}_{\rho} \Gamma^{\sigma}_{[\nu \mu]} \tau_{\sigma} + \mathrm{cyclic}(\mu,\nu,\rho)
	   &=& \nabla_{[\mu} \mathbbm{E}_{\nu]} \tau_{\rho} + \mathrm{cyclic}(\mu,\nu,\rho) \; , 
	\end{eqnarray}
where we have employed the metric compatibility of $\tau_{\mu}$ i.e. $\nabla_{\mu} \tau_{\nu} = 0$. Contractions of Eq.~\eqref{Eq:BianchiIdentityIII} yield additional constraints we can use for the construction of the generating functional. As Eq. \eqref{Eq:BianchiIdentityIII} is completely antisymmetric, the only structure we can contract it with to yield a non-trivial result is $u^{\mu} \nu^{\nu} h^{\rho \sigma} \mathbbm{E}_{\sigma}$. Consequently,
\begin{eqnarray}
		\label{Eq:Torsion3}
			-  \nu^{\nu} h^{\rho \alpha} \mathbbm{E}_{\alpha} \Gamma_{[\nu \rho]}^{\sigma} \mathbbm{E}_{\sigma} 
		 &=& \nu^{\nu} h^{\rho \sigma} \mathbbm{E}_{\sigma}  \nabla_{\left[ \nu \right.} \mathbbm{E}_{\left. \rho \right]} 
	     	        -  u^{\mu} h^{\rho \sigma} \mathbbm{E}_{\sigma} \nabla_{\left[ \rho \right.} \mathbbm{E}_{\left. \mu \right]}  + \mathbbm{E}^2 u^{\mu} \nu^{\nu} \Gamma_{[\nu \mu]}^{\sigma} \tau_{\sigma}  \nonumber \\
		  &\;& -  u^{\mu} h^{\rho \sigma} \mathbbm{E}_{\sigma} \Gamma_{[\rho \mu]}^{\alpha} \mathbbm{E}_{\alpha}
		  	 + (\mathbbm{E} \cdot u) \nu^{\nu} h^{\rho \sigma} \mathbbm{E}_{\sigma} \Gamma_{[\rho \nu]}^{\alpha} \tau_{\alpha} \; ,
	\end{eqnarray} 
which again allows us to eliminate torsion terms. Meanwhile in FSCC the Bianchi identity \eqref{Eq:BianchiIdentityII} reduces to
\begin{equation}
    \label{Eq:FlatspaceBianchi}
    \partial_j\mathbbm{E}_i-\partial_i\mathbbm{E}_j=0 \; . 
\end{equation}
}

{\ Finally, we impose hydrostatic constraints on $\nu^\mu$ and $h_{\mu\nu}$. The Killing constraint evaluated on $\nu^\mu$ reads
	\begin{eqnarray}
		\label{Eq:NuKillingCondition}
	 	0 &=& u^{\nu} \partial_{\nu} \nu^{\mu} - \nu^{\nu} \partial_{\nu} u^{\mu} + u^{\mu} \nu^{\nu} \frac{\partial_{\nu} T}{T} \; ,  
	\end{eqnarray}
or in explicitly covariantised form
	\begin{eqnarray}
		\label{Eq:Torsion4} 
			- 2 u^{\nu} \Gamma_{[\nu \rho]}^{\mu} \nu^{\rho}
		&=& u^{\nu} \nabla_{\nu} \nu^{\mu} - \nu^{\nu} \nabla_{\nu} u^{\mu} + u^{\mu} \nu^{\nu} \frac{\nabla_{\nu} T}{T} \; .  
	\end{eqnarray}
In flat space, \eqref{Eq:NuKillingCondition} reduces to the following independent constraint
\begin{equation}
    \label{Eq:noaccn}
    \partial_tv^i=0 \; ,
\end{equation}
which generalizes the second equation in \eqref{Eq:Electricfieldconstraints}.}

{\ For our last constraint, we consider $h_{\mu \nu}$, i.e. \eqref{Eq:Killingh}. The resultant expression evaluates to
\begin{equation}
    \mathcal{L}_uh_{\mu\nu}-\frac{u^\rho}{T}h_{\rho\nu}\partial_\mu T-\frac{u^\rho}{T}h_{\rho\mu}\partial_\nu T=0
\end{equation}
or 
	\begin{eqnarray}
		\label{Eq:Torsion5}
		2 u^{\sigma} \Gamma_{[\sigma \mu]}^{\rho} h_{\rho \nu} + 2 u^{\sigma} \Gamma_{[\sigma \nu]}^{\rho} h_{\mu \rho}
		&=& - \nabla_{\mu} u^{\sigma} h_{\sigma \nu} - \nabla_{\nu} u^{\sigma} h_{\mu \sigma} 
			+ \frac{u^\rho}{T}h_{\rho\nu}\nabla_\mu T \nonumber \\
		&\;& + \frac{u^\rho}{T}h_{\rho\mu}\nabla_\nu T \; . 
	\end{eqnarray}
Again, in flat space we find the independent constraint
\begin{equation}
   \label{Eq:Squarevconstraints}
    \partial_{i} v_{j} + \partial_{j} v_{i} = 0 \; . 
\end{equation}
where we have employed $\partial_{\mu} T =0$. Equation \eqref{Eq:Squarevconstraints} is nothing more than the boost agnostic version of the usual Killing condition on the velocity found in relativistic hydrostatic fluids \cite{PhysRevLett.109.101601}. }

{\ }

{\noindent \textbf{Summary of the hydrostatic conditions in flat spacetime} - For convenience, we  list here all the independent hydrostatic conditions in the FSCC limit. Firstly, any scalar quantities must satisfy $\mathcal{L}_{u}(\dots)=(\partial_t+v^i\partial_i)(\dots)=0$. Then, additionally, we have shown that:
\begin{align}
    \label{Eq:OrderZeroConstraints}
    &\partial_\mu T=0  \; , & &  \partial_t v^i = 0  \; ,   & & \partial_i v_j + \partial_j v_i =0  \; , \qquad \nonumber \\
    & \partial_i\mathbbm{E}_j-\partial_j\mathbbm{E}_i=0 \; , &   &  \partial_{t} \mathbbm{E}_{i} + v^{j} \partial_{j} \mathbbm{E}_{i} +  \mathbbm{E}_{j} \partial_{i} v^{j} = 0  \; , & &
    \end{align}
and
\begin{eqnarray}
	\label{Eq:ElectricFieldBalance}
	\mathbbm{E}_{i} - \partial_{i} \mu &=& 0 \; . 
\end{eqnarray}
In the following sections, we use these constraints to write down the most general  hydrostatic effective action for our fluids. In appendix \ref{appendix:Conserved} we also explicitly check that the resulting fluids identically satisfy the hydrodynamic equations of motion in the hydrostatic regime, thus explicitly demonstrating that a zeroth order electric fields is compatible with hydrostatic equilibrium.}

\subsection{Hydrostatic constitutive relations}\label{subsec:HSConRelations}

{\noindent Now that we have all the hydrostatic conditions \eqref{Eq:OrderZeroConstraints}, we can apply the generating functional method. To do so we must construct a scalar functional $W$ out of the independent scalars built out of our thermodynamic quantities and their derivatives. In particular, we build $W$ out of the scalars presented in table \ref{Tab:OrderZeroQuantities}.

 \begin{table}[!h]
	\centering
	\begin{tabular}{|c||c|c|} \hline
		& Elementary & Composite \\ \hline 
		\textbf{Scalars:} & $T , \; \mu$ & $ h_{\mu \nu} u^{\mu} u^{\nu} , \; h^{\mu \nu} \mathbbm{E}_{\mu} \mathbbm{E}_{\mu}, \; \mathbbm{E}_{\mu} u^{\mu}$ \\ \hline
		\textbf{One-forms:} & $\tau_{\mu}, \;  \mathbbm{E}_{\mu}, \; \partial_{\mu} \mu $ & $ h_{\mu \nu} u^{\mu} $\\ \hline
		\textbf{Vectors:} & $\nu^{\mu}, \; u^{\mu}$ & $ h^{\mu \nu} \mathbbm{E}_{\nu}\; , h^{\mu \nu} \partial_{\mu} \mu$ \\ \hline 
		\textbf{Covariant 2-tensors:} & $h_{\mu \nu}$ & $\tau_{\mu} \tau_{\nu}, \; \tau_{\mu} \mathbbm{E}_{\nu}, \; \tau_{\mu} \partial_{\nu} \mu \ldots$ \\ \hline
		\textbf{Contravariant 2-tensors:} & $h^{\mu \nu}$ & $\nu^{\mu} \nu^{\nu}, \; \nu^{\mu} u^{\nu}, \; u^{\mu} u^{\nu}$ \\ \hline
	\end{tabular}
	\caption{Independent in principle scalars, vectors and tensors at zeroth order in derivatives. We do not include the gauge field as it is not gauge invariant and can only enter the generating functional as an antisymmetrised derivative (electric field) or Wilson loop (chemical potential). The generic terms that can appear in the order one generating functional are given by taking the covariant derivative of the elementary scalars and contracting the result with the elementary \& composite quantities to give scalars.}
	\label{Tab:OrderZeroQuantities}
\end{table} 

 Restricting ourselves to zeroth order in derivatives, we define the zeroth order generating functional $W_{(0)}$ in the presence of the sources  ($\tau_{\mu}$, $h_{\mu \nu}$, $A_{\mu}$ and $F_{\mu \nu}$) to be
\begin{align}
    \label{Eq:ConstitutiveGenerator}
    W_{(0)}[\tau,h,A,F]&=\int d^{d+1}x\; e \; P(T,\mu, h^{\mu \nu} \mathbbm{E}_{\mu} \mathbbm{E}_{\nu}, h_{\mu \nu} u^{\mu} u^{\nu} , \mathbbm{E}_{\mu} u^{\mu})
\\
&\equiv \int d^{d+1}x\; e\; P\left(T,\mu,\vec{\mathbbm{E}}^2,\vec{v}^2,\vec{v} \cdot \vec{\mathbbm{E}} \right)\nonumber
\end{align}
with $e = \mathrm{det}(\tau_\mu,e^a_\mu)$. By varying the above generating functional with respect to the background sources we define the following one-point functions
\begin{equation}
    \label{Eq:Defof1pt}
    \delta W_{(0)}[\tau,h,A,F]=\int d^{d+1}x \; e\left(-T^\mu\delta\tau_\mu+\frac{1}{2}T^{\mu\nu}\delta h_{\mu\nu}+J^\mu\delta A_\mu + \frac{1}{2}  M^{\mu \nu} \delta F_{\mu \nu}\right)~,
\end{equation}
where $T^\mu$ is the energy current, $T^{\mu\nu}$ is the stress-momentum tensor, $J^\mu$ the U(1) charge current and $M^{\mu\nu}$ the magnetization density tensor. Out of the energy current and stress-momentum tensor, we can assemble a ``stress-energy-momentum'' (or SEM) tensor as the combination
\begin{equation}
    \label{Eq:SEMtensor}
    T\indices{^\mu_\nu}=-T^\mu\tau_\nu+T^{\mu\rho}h_{\rho\nu} \; .
\end{equation}
Notice that $T^\mu_{\ \nu}$ is still symmetric in its spatial indices since we are assuming rotational invariance of the microscopic theory. Additionally, we can decompose $M^{\mu\nu}$ as
	\begin{eqnarray}
		M^{\mu \nu} &=& \nu^{\mu} \mathbbm{P}^{\nu} - \nu^{\nu} \mathbbm{P}^{\mu} + h^{\mu \rho} h_{\rho \alpha} h^{\nu \sigma} h_{\sigma \beta} M^{\alpha \beta}
	\end{eqnarray}
where $\mathbbm{P}^{\mu}$ is the polarization vector and the final term can be further decomposed into magnetisation scalars, vectors, tensors etc. In what follows, only $\mathbbm{P}^{\mu}$ will be non-zero since we are not considering magnetic fields in our setup.}

{\ Some parenthetical comments about \eqref{Eq:ConstitutiveGenerator} and \eqref{Eq:Defof1pt} are in order. Firstly, in \eqref{Eq:Defof1pt}, the field strength is not independent of the gauge field $A_{\mu}$ and thus there is some overcounting of degrees of freedom. Nevertheless, separating these quantities in this manner is quite useful as $F_{\mu \nu}$ consists only of gauge invariant degrees of freedom while $A_{\mu}$ includes non-local terms like the chemical potential.}

{\ Secondly, we should make note of the unusual (to the literature) choice for the discrete symmetries of our setup. Our theory does not have parity-breaking parameters, so all the scalars in \eqref{Eq:ConstitutiveGenerator} are $\mathcal{P}$-even. However the scalar $\vec{\mathbb{E}}\cdot\vec{v}$ is $\mathcal{T}$-odd, which means that the thermodynamics, generally, will not be time-reversal covariant. By this we mean that the pressure need not have a definite sign under time-reversal. This is also true for \eqref{Eq:VelocityRelation} if we make generic choices for the $\Omega$ terms and in such cases one should not expect the Onsager relations to hold.}

{\ With that said, requiring $W_{(0)}$ to be diffeomorphism and gauge invariant yields (non-)conservation equations for the SEM and the charge current
	\begin{subequations}
	\begin{eqnarray}
	   \label{Eq:SEMconscurved}
	   e^{-1} \partial_{\mu} \left( e T\indices{^\mu_\rho} \right) + T^{\mu} \partial_{\rho} \tau_{\mu} - \frac{1}{2} T^{\mu \nu} \partial_{\rho} h_{\mu \nu} &=& F_{\rho \mu} J^{\mu} \; , \\
	   \label{Eq:Chargeconscurved}
	   e^{-1} \partial_{\mu} \left( e J^{\mu} \right) &=& 0 \; ,
	\end{eqnarray}
where we have employed \eqref{Eq:Defof1pt} and \eqref{Eq:SEMtensor}. In addition we have the following conservation equation for the polarisation
	\begin{eqnarray}
		   e^{-1} \partial_{\mu} \partial_\nu\left(2e\nu^{[\mu}\mathbbm{P}^{\nu]}\right) &=& 0 \; .\label{Eq:PolarizationConservation}
	\end{eqnarray}
	\end{subequations}
These equations are covariant which can be shown by replacing the partial derivative by a suitable covariant derivative as discussed in \cite{deBoer:2020xlc}.}

{\ Combining the explicit form of $W_{(0)}$ \eqref{Eq:ConstitutiveGenerator} and the definition of the 1-point functions \eqref{Eq:Defof1pt}, we can derive the constitutive relations for our fluid. In doing so, we use the following variations with respect to the background sources
\begin{subequations}
\begin{eqnarray}
    \delta T&=&-Tu^\mu\delta\tau_\mu \; , \\
    \delta u^\mu&=&-u^\mu u^\nu\delta\tau_\nu \; , \\
    \delta\mathbbm{E}_\mu&=&\mathbbm{E}_\mu \nu^\rho\delta\tau_\rho - \tau_{\mu} \mathbbm{E}_{\nu}h^{\nu(\rho}\nu^{\sigma)}\delta h_{\rho\sigma}-(\partial_\mu\delta A_\nu-\partial_\nu\delta A_\mu)\nu^\nu \; , \\
    \delta h^{\mu\nu}&=&-h^{\mu\rho}\delta h_{\rho\sigma}h^{\sigma\nu}+2\nu^{(\mu}h^{\nu)\rho} \delta\tau_\rho \; , \\
    \delta \nu^\mu&=& \nu^\mu \nu^\nu\delta\tau_\nu-h^{\mu(\nu}\nu^{\rho)}\delta h_{\nu\rho} \; , \\
    \delta\mu&=&-\mu u^\mu\delta\tau_\mu+u^\mu\delta A_\mu \; , \\
    \delta e&=&e\left(-\nu^\mu\delta\tau_\mu+\frac{1}{2}h^{\mu\nu}\delta h_{\mu\nu}\right) \; , 
\end{eqnarray}
\end{subequations}
and recall the identities
\begin{subequations}
\begin{eqnarray}
    \nu^\mu&=-u^\mu+h^{\mu\rho}h_{\rho\nu}u^\nu \; , \\
    F_{\mu\nu}&=\mathbbm{E}_\mu\tau_\nu-\mathbbm{E}_\nu\tau_\mu \; .
\end{eqnarray}
\end{subequations}
We find the following expressions for the constitutive relations at zeroth order in derivatives
\begin{align}
    T^\mu&=\varepsilon u^\mu+\left(P-\mathbbm{P}^{\sigma} \mathbbm{E}_{\sigma} \right)h^{\mu\rho}h_{\rho\nu}u^\nu \; , \\
    T^{\mu\nu}&=P h^{\mu\nu}+\rho_{\mathrm{m}}u^\mu u^\nu-\kappa_{\mathbbm{E}}h^{\mu\rho}\mathbbm{E}_\rho h^{\nu\sigma}\mathbbm{E}_\sigma- 2\beta_{\mathbbm{P}} \mathbbm{E}_\rho h^{\rho(\mu}\nu^{\nu)} \; 
\end{align}
and
\begin{subequations}
\label{Eq:ConstitutiveRelations0unflat}
\begin{eqnarray}
    T^\mu_{\ \nu}&=&-\varepsilon u^\mu\tau_\nu-\left(P -\mathbbm{P}^{\sigma} \mathbbm{E}_{\sigma}\right)h^{\mu\rho}h_{\rho\sigma}u^\sigma\tau_\nu+P h^{\mu\rho}h_{\rho\nu}+\rho_{\mathrm{m}}u^\mu u^\rho h_{\rho\nu}\nonumber\\
    &\;& -\kappa_{\mathbbm{E}}\mathbbm{E}_\alpha\mathbbm{E}_\beta h^{\alpha\mu}h^{\beta\rho}h_{\rho\nu}-\beta_{\mathbbm{P}} \mathbbm{E}_\alpha h^{\alpha\rho}\nu^{\mu}h_{\rho\nu}\\
    J^\mu&=&n u^\mu+\frac{1}{e}\partial_\nu\left(2e\nu^{[\mu}\mathbbm{P}^{\nu]}\right)~.\label{Eq:ConstitutiveRelations0Current}
\end{eqnarray}
\end{subequations}
In \eqref{Eq:ConstitutiveRelations0unflat}, $P$ is the fluid pressure, $\varepsilon$ its energy density, $\rho_{\mathrm{m}}$ its momentum density\footnote{$\rho_m$ is called kinetic mass density in \cite{deBoer:2017ing}.} and $n$ its charge/number density. In thermodynamic equilibrium, they are defined as
\begin{align}
\label{Eq:Densities}
n =&\left( \frac{\partial P}{\partial \mu} \right) \; ,~~\rho_{\mathrm{m}} = 2 \left( \frac{\partial P}{\partial \vec{v}^2} \right) \; ,~~  s = \left( \frac{\partial P}{\partial T} \right) \; , ~~ \\
& \beta_{\mathbbm{P}} = \left( \frac{\partial P}{\partial (\vec{\mathbbm{E}} \cdot \vec{v})} \right) \; , ~~
\kappa_{\mathbbm{E}} =  2 \left( \frac{\partial P}{\partial \vec{\mathbbm{E}}^2} \right) \; ,  
\end{align}
where $s$ is the entropy density of the fluid.  Together they satisfy
\begin{align}
\varepsilon+P&=sT+\mu n+\rho_{\mathrm{m}} \vec{v} ^2+\kappa_{\mathbbm{E}} \vec{\mathbbm{E}}^2+2\beta_{\mathbbm{P}} \vec{\mathbbm{E}} \cdot \vec{v}\nonumber~.
\end{align}
The parameters $\kappa_{\mathbbm{E}}$ and $\beta_{\mathbbm{P}}$ do not appear in the hydrodynamics literature. To understand their physical meaning, we note that we can write the momentum and polarization of the fluid as 
\begin{align}
\label{Eq:polarisationdef}
		 \vec{\mathbbm{P}} =& \left( \frac{\partial P}{\partial \vec{\mathbbm{E}}} \right) = \kappa_{\mathbbm{E}} \vec{\mathbbm{E}} + \beta_{\mathbbm{P}} \vec{v} \; , \\
		 \label{Eq:momentumdef}
		\vec{P} =& \left( \frac{\partial P}{\partial \vec{v}} \right)  = \rho_{\mathrm{m}} \vec{v} + \beta_{\mathbbm{P}} \vec{\mathbbm{E}}   \; . 
 \end{align}
Notice that in the present case the system can have a non-zero polarization even when there are no external electromagnetic fields, purely due to the velocity. This effect can be traced back to the fact that our thermodynamics is $\mathcal{T}$-odd. If we instead insist in having a theory that is time-reversal invariant (so the pressure depends on even powers of $\vec{\mathbb{E}}\cdot\vec{v}$) then $\beta_\mathbb{P}$ would depend on odd powers of $\vec{\mathbb{E}}\cdot\vec{v}$ and disappear at zero electric field. This is exactly the same effect observed in \cite{Kovtun:2016lfw} for a relativistic fluid, in which the scalar $E\cdot B$ breaks time-reversal and thus the system can have a non-zero polarization at zero electric field given by the magnetic field alone.}
	
{\ In terms of the momentum and density of the fluid, we can re-express the definition of the energy density $\varepsilon$ \eqref{Eq:Densities} as
	\begin{eqnarray}
	\label{Eq:GibbsDuhem}
	 \varepsilon + P &=&sT + \vec{\mathbbm{E}}\cdot \vec{\mathbbm{P}} + \vec{v} \cdot \vec{P} + n \mu  \; .  
	\end{eqnarray}
	Subsequently we have that
	\begin{eqnarray}
	\label{Eq:InfinitesimalPressure}
	 \mathrm{d} P &=& s \mathrm{d} T + n \mathrm{d} \mu + P_{i} \mathrm{d} v^{i} + \mathbbm{P}^{i} \mathrm{d} \mathbbm{E}_{i} \; . 
	\end{eqnarray}
These are nothing more than the Gibbs-Duhem relation and 1st law of thermodynamics for our fluid, respectively.

Some comments regarding our constitutive relations \eqref{Eq:ConstitutiveRelations0unflat} are in order. First, we note that the spatial momentum $\vec{P}$ in our formalism is an observable quantity with a definite value. This contrasts the boost invariant case, where we can generally choose a frame for which $\vec{P}$ vanishes in global thermodynamic equilibrium. Second, the derivative term in the current is the polarization contribution to the bound current \cite{Kovtun:2016lfw}. Equation \eqref{Eq:ConstitutiveRelations0Current} together with \eqref{Eq:PolarizationConservation} then imply that the free charges (parametrised by $n$) and the bound charges, i.e. the polarization, are separately conserved. Moreover, there is a subtle point worth noticing about the constitutive relations in \eqref{Eq:ConstitutiveRelations0unflat}. Typically the velocity field $u^{\mu}$ is an eigenvector of the SEM tensor at lowest order in derivatives. Imposing this at higher orders constitutes the so called Landau-frame choice. However, for our fluid
	\begin{eqnarray}
	 	 T\indices{^{\mu}_{\nu}} u^{\nu} 
	 &=&	 - \left( \varepsilon - \rho_{\mathrm{m}} u^2 - \mathbbm{P} \cdot \mathbbm{E} \right) u^{\mu} + \left( \mathbbm{P} \cdot \mathbbm{E} - \beta_{\mathbbm{P}}(\mathbbm{E} \cdot u)  \right) \nu^{\mu} \nonumber \\
	 &\;& - \kappa_{\mathbbm{E}} (\mathbbm{E} \cdot u) h^{\mu \nu} \mathbbm{E}_{\nu}  \; .
	\end{eqnarray}
In the absence of the electric field one finds that the fluid velocity is indeed an eigenvector of the SEM tensor with an eigenvalue that comprises the energy density minus the kinetic energy density. Yet, in the presence of the electric field this is not the case and the naive Landau frame does not exist for our fluid.\footnote{The absence of Landau frame already happens in relativistic hydrodynamics, see \cite{Kovtun:2016lfw}.} Instead a particular thermodynamic frame, given by our definition of the hydrodynamic fields \eqref{Eq:tempdef} and \eqref{Eq:TimeComponentu}, is well-defined and we shall use this frame choice throughout the rest of the paper.}

{\ Finally, our constitutive relations \eqref{Eq:ConstitutiveRelations0unflat} reduce in the FSCC limit to 	\begin{subequations}
	\label{Eq:ConstitutiveRelations0}
	\begin{eqnarray}
    T\indices{^0_0}&=&-\varepsilon \; , \\
    T\indices{^0_i}&=&\rho_{\mathrm{m}}v_i+\beta_{\mathbbm{P}} \mathbbm{E}_i=P_i \; , \\
    T\indices{^i_0}&=&-\left(\varepsilon+P-\vec{\mathbbm{P}} \cdot \vec{\mathbbm{E}} \right)v^i \; , \\
    T\indices{^i_j}&=&P \delta^i_j+\rho_{\mathrm{m}}v^iv_j-\kappa_{\mathbbm{E}}\mathbbm{E}^i\mathbbm{E}_j \; , \\
    J^0&=&n-\partial_j\mathbbm{P}^j \; , \\
    \label{Eq:SpatialChargeCurrent}
    J^i&=& nv^i+\partial_t\mathbbm{P}^i \; . 
\end{eqnarray}
	\end{subequations}
}

{\ Notice that independent constant electric and velocity fields are solutions of the hydrodynamic equations of motion \eqref{Eq:OrderZeroConstraints} and \eqref{Eq:ElectricFieldBalance}. As a result, enforcing a Drude-like constraint such as  \eqref{Eq:VelocityRelation} is not necessary. However, the electric field cannot take arbitrary values---as one would expect in experiments---but must be carefully balanced by a non-trivial profile for the chemical potential for the system to remain in hydrostatic equilibrium. Consequently, an electric field driven flow and current seem impossible to generate in hydrostatic equilibrium.
To circumvent this issue, one could attempt to impose \eqref{Eq:VelocityRelation} on the boost agnostic hydrodynamics directly, without including additional relaxation terms. This is an unphysical route to follow for the following reasons.
 
 First, within the hydrodynamic theory we have already written down, the Drude constraint \eqref{Eq:VelocityRelation} can only arise as a particular solution of the equations of motion. Thus, enforcing \eqref{Eq:VelocityRelation} does not extend the theory beyond hydrostatic equilibrium, but rather constricts it to a narrow subset of phase space. Further, compatibility between the Drude and hydrostatic constraints yields additional constraints on $\Omega_{\mathbbm{E}}$ and $\Omega_{\mu}$. For example, we can use \eqref{Eq:ElectricFieldBalance} to rewrite \eqref{Eq:VelocityRelation}  as
	\begin{eqnarray}
		\label{Eq:Simplifiedflow}
		v^{i} &=& \tilde{\Omega}_{\mathbbm{E}}(T,\mu,\vec{v}^2,\vec{\mathbbm{E}}^2,\vec{v} \cdot \vec{\mathbbm{E}}) \mathbbm{E}^{i} \; , \qquad \tilde{\Omega}_{\mathbbm{E}} = \Omega_{\mathbbm{E}} + \Omega_{\mu} \; . 
	\end{eqnarray}
Substituting this expression into \eqref{Eq:OrderZeroConstraints} one finds constraints that lead to the thermodynamic derivatives of $\tilde{\Omega}_{\mathbbm{E}}$ not being free variables (and thus neither is $\tilde{\Omega}_{\mathbbm{E}}$ free).  If we  assume this applies to every system, then we have derived an extremely strong constraint on the kinds of transport one can encounter in nature. For example, using one of the stationarity constraints \eqref{Eq:OrderZeroConstraints} with \eqref{Eq:Simplifiedflow} leads to:
	\begin{eqnarray}
		\label{Eq:symvelocityconst}
		0 &=& \partial_{(i} v_{j)} 
		     =  \frac{\partial \tilde{\Omega}_{\mathbbm{E}}}{\partial \mu} \mathbbm{E}_{(i} \mathbbm{E}_{j)}
		   	  +  \frac{\partial \tilde{\Omega}_{\mathbbm{E}}}{\partial \vec{\mathbbm{E}}^2}  \mathbbm{E}_{(i} \partial_{j)} \vec{\mathbbm{E}}^2 +  \frac{\partial \tilde{\Omega}_{\mathbbm{E}}}{\partial \vec{v}^2}  \mathbbm{E}_{(i} \partial_{j)} \vec{v}^2 \nonumber \\
		   &\;& \hphantom{\partial_{(i} v_{j)}  =} +  \frac{\partial \tilde{\Omega}_{\mathbbm{E}}}{\partial (\vec{\mathbbm{E}} \cdot \vec{v})}  \mathbbm{E}_{(i} \partial_{j)} (\vec{\mathbbm{E}} \cdot \vec{v}) + \tilde{\Omega}_{\mathbbm{E}} \partial_{(i} \mathbbm{E}_{j)} \; .
	\end{eqnarray}
Comparing tensor structures we see the first term implies that $\tilde{\Omega}_{\mathbbm{E}}$ is independent of the chemical potential. In addition, it is not hard to see that at lowest order in derivatives $\tilde{\Omega}_{\mathbbm{E}}$ plays the role of the DC conductivity $\sigma_{\rm DC}$
	\begin{eqnarray}
		\vec{J} = n \tilde{\Omega}_{\mathbbm{E}} \vec{\mathbbm{E}} \; , \qquad \sigma_{\mathrm{DC}} = n \tilde{\Omega}_{\mathbbm{E}} \; .
	\end{eqnarray}
Unfortunately we cannot think of this term as some kind of incoherent conductivity \cite{Davison:2015bea} as the latter is generically non-zero at vanishing chemical potential (where the charge density is also expected to vanish). Therefore, the DC conductivity depends on the chemical potential entirely through the number density $n$. While we have not been able to rule out such a situation from first principles, we do not find this scenario credible and henceforth we shall assume that $\tilde{\Omega}_{\mathbbm{E}}$ is not constrained, but ultimately it will be the hydrostatic constraint \eqref{Eq:ElectricFieldBalance} that needs to be modified by the presence of relaxation terms. In this case $\Omega_{\mu} \neq 0$ generically, but $\Omega_{\mu} = 0$ can be reached as a special limit by tuning the relaxation terms appropriately.}

{\ We proceed in the next subsection by introducing relaxation terms and deriving how  the hydrostatic constraint \eqref{Eq:ElectricFieldBalance} must be modified.} 

\subsection{Relaxation at zeroth order}\label{subsec:0thOrderRelaxation}

{\noindent In the hydrodynamic formulation we have developed, $\vec{v}$ and $\vec{\mathbbm{E}}$ are independent variables. In the hydrostatic limit they can assume any functional form, consistent with \eqref{Eq:OrderZeroConstraints}, independently from one another and our hydrodynamic equations will still be satisfied. In particular, we do not have to obey the constraint \eqref{Eq:VelocityRelation}. To move away from the hydrostatic result, we follow our intuition from the Drude model and relax the equations of motion with decay terms.}

{\ 
We expect that to be consistent with the principle of effective field theory, the relaxation terms should be written in terms of the effective operators in the theory. At order zero in derivatives these operators can only be the hydrostatic ones such as the energy density, number density, the momentum etc. These form one class of relaxations---those that are non-zero at stationarity---and as we shall see such terms can actually lead to modifications of the hydrostatic constraints (e.g. \eqref{Eq:OrderZeroConstraints}). At order one, an additional class of relaxation terms consisting of non-hydrostatic operators is possible.}

{\ With these points in mind, we add relaxation in the natural way familiar from Lagrangian mechanics; we derive the conservative parts of the equations of motions from $W_{(0)}$, i.e. Eq.s \eqref{Eq:Chargeconscurved}, through a variational principle and subsequently add in the non-conservative forces by hand and check for consistency. We carry out this analysis only in the FSCC limit, where our relaxed conservation equations take the following form
	\begin{subequations}
	\label{Eq:Relaxed0thOrder}
	\begin{eqnarray}
			   \label{Eq:Energyrelaxationgeneric}
		0 &=& \partial_{t} \varepsilon + \partial_{i} J^{i}_{\varepsilon} - \mathbbm{E}_{i} J^{i} 
			  + \hat{\Gamma}_{\varepsilon} \; , \\
			   \label{Eq:Momentumrelaxationgeneric}
		0 &=& \partial_{t}  P_{i} + \partial_{j} T^{ij} - n J^{i} + \hat{\Gamma}_{\vec{P}}^{i} \; , \\
			   \label{Eq:Chargerelaxationgeneric}
		0 &=& \partial_{t} n + \partial_{i} J^{i} \; . 
	\end{eqnarray}
	\end{subequations}
The relaxation terms displayed in \eqref{Eq:Energyrelaxationgeneric}-\eqref{Eq:Chargerelaxationgeneric} are the most generic ones we can add to the constitutive relations at lowest order in derivatives. In particular, momentum relaxation can only include $(P_{i}, \mathbbm{P}^{i})$  because we are assuming that $\partial^{i} \mu$ (which, as a reminder, can be order zero in derivatives) can still be expressed in terms of the electric field.\footnote{To be justified shortly.}}

{\ Our goal now is to determine the relaxation terms such that the equations of motion are identically satisfied. The conservation equations split into 0th and 1st order pieces, which need to be satisfied independently for our derivative expansion to be consistent. Focusing for the moment only on the zeroth order terms, we identify
   \begin{eqnarray}
   \label{Eq:RelaxIntro}
     \hat{\Gamma}_{\vec{P}}^{i} = \Gamma_{\vec{\mathbbm{P}}} \mathbbm{P}^{i} + \Gamma_{\vec{P}} P^{i} &+& \mathcal{O}(\partial) \; , \qquad  \hat{\Gamma}_{\varepsilon} = \Gamma_{\varepsilon} + \mathcal{O}(\partial)  \; .
    \end{eqnarray} 
One may be tempted to add terms of the form
	\begin{eqnarray}
		\left(\Gamma_{\vec{P}}\right)\indices{_i^j} P_{j} \; , \qquad \left(\Gamma_{\vec{\mathbbm{P}}}\right)\indices{_i^j} \mathbbm{P}_{j} \; , 
	\end{eqnarray}
where the relaxation terms are matrices to $\hat{\Gamma}^{i}_{\vec{P}}$. However, requiring the relaxation rates be expressible in terms of $(P_{i}, \mathbbm{P}^{i})$---as is appropriate for an effective theory---we find that
	\begin{eqnarray}
		\left(\Gamma_{\vec{P}}\right)\indices{_i^j} P_{j} &=& \left( \alpha_{1} P_{i} P^{j} + \alpha_{2} \mathbbm{P}_{i} P^{j} + \alpha_{3} \mathbbm{P}^{j} P_{i} + \alpha_{4} \mathbbm{P}_{i} \mathbbm{P}^{j}  + \alpha_{5} \delta_i^{j} \right) P_{j} \nonumber \\
											&=& \left( \alpha_{1} \vec{P}^2 + \alpha_{2} \vec{\mathbbm{P}} \cdot \vec{P} + \alpha_{5} \right) P_{i} + \left( \alpha_{3} \vec{P} \cdot \vec{\mathbbm{P}} + \alpha_{4} \vec{\mathbbm{P}}^2 \right) \mathbbm{P}_{i} \; ,
	\end{eqnarray}
i.e. we ca express the relaxation rates as a linear combination of  $P^{i}$ and $\mathbbm{P}^{i}$ and, thus, our choice of relaxation terms in \eqref{Eq:RelaxIntro} is the most general one.}

{\ Subsequently, the resultant (non-)conservation equations of motion take the form %
    \begin{subequations}
    \label{Eq:ViolatingConservation}
    \begin{eqnarray}
     \label{Eq:EnergyDissipationDef}
     n v^{i} \left( \mathbbm{E}_{i}  -   \partial_{i} \mu \right)  &=& \Gamma_{\varepsilon} + \mathcal{O}(\partial) \; , \\
     \label{Eq:MomentumDissipationDef}
     n \left( \mathbbm{E}^{i} - \partial^{i} \mu \right) &=&  \Gamma_{\vec{\mathbbm{P}}} \mathbbm{P}^{i} + \Gamma_{\vec{P}} P^{i} + \mathcal{O}(\partial) \; .
     \end{eqnarray}
    \end{subequations}
     We observe that hydrostatic constraints \eqref{Eq:OrderZeroConstraints} no longer identically satisfy \eqref{Eq:ViolatingConservation} and we must re-examine the equations of motion.  As a crosscheck, we note that in the absence of relaxation terms, we still need to impose the same constraint \eqref{Eq:ViolatingConservation} as before, namely:
	\begin{eqnarray}
		\label{Eq:Usualconstraint}
		\mathbbm{E}^{i} - \partial^{i} \mu &=& 0 \; .
 	\end{eqnarray}
}

{\ Let us now analyse various generic situations in the presence of relaxation without imposing \eqref{Eq:VelocityRelation}. If we maintain \eqref{Eq:Usualconstraint} while retaining relaxation terms we find we must require
	\begin{subequations}
	\label{Eq:Nomodificationstationarity}
	\begin{eqnarray}
	\label{Eq:NoModMomentum}\Gamma_{\vec{\mathbbm{P}}} \mathbbm{P}^{i} + \Gamma_{\vec{P}} P^{i} &=& 0 \; , 
	\\ 
	\label{Eq:NoModEnergy}  \Gamma_{\varepsilon} = 0 \; .
	\end{eqnarray}
	\end{subequations}
Re-expressing \eqref{Eq:Nomodificationstationarity} in terms of $\vec{v}$ and $\vec{\mathbb{E}}$ we find 
	\begin{eqnarray}
		\label{Eq:VelocityNomodification}
		v^{i} &=& - \left( \frac{\kappa_{\mathbbm{E}} \Gamma_{\vec{\mathbbm{P}}} + \beta_{\mathbbm{P}} \Gamma_{\vec{P}}}{ \beta_{\mathbbm{P}} \Gamma_{\vec{\mathbbm{P}}} + \rho_{\mathrm{m}} \Gamma_{\vec{P}}} \right)  \mathbbm{E}^{i} \; .
	\end{eqnarray}
We see that Eq.~\eqref{Eq:VelocityNomodification} provides us with the Drude constraint generalized in the hydrodynamic setting. Comparing \eqref{Eq:VelocityNomodification} with \eqref{Eq:VelocityRelation} we see that $\tilde{\Omega}_{\mathbbm{E}}$, defined in \eqref{Eq:Simplifiedflow}, is given by
	\begin{eqnarray}
		\tilde{\Omega}_{\mathbbm{E}} = - \left( \frac{\kappa_{\mathbbm{E}} \Gamma_{\vec{\mathbbm{P}}} + \beta_{\mathbbm{P}} \Gamma_{\vec{P}}}{\beta_{\mathbbm{P}} \Gamma_{\vec{\mathbbm{P}}} + \Gamma_{\vec{P}} \kappa_{\mathbbm{E}}} \right) \; .
	\end{eqnarray}
Subsequently one finds the constraints on the DC conductivity discussed near the end of subsection \ref{subsec:HSConRelations}.}

{\ The conditions \eqref{Eq:Nomodificationstationarity} are a limit of our generic results where the energy and momentum relaxations are unconstrained. Returning to solving the full expressions presented in \eqref{Eq:ViolatingConservation}, we now allow ourselves to violate \eqref{Eq:Usualconstraint}. Firstly, we note that equation \eqref{Eq:EnergyDissipationDef} is not independent of \eqref{Eq:MomentumDissipationDef} on-shell as can be seen by contracting \eqref{Eq:MomentumDissipationDef}  with the spatial fluid velocity. In particular, we can identify $\Gamma_{\varepsilon}$ in terms of $\Gamma_{\vec{P}}$ and $\Gamma_{\vec{\mathbbm{P}}}$,
	\begin{eqnarray}
	\label{Eq:IdentificationofGammaE}
	 \Gamma_{\varepsilon} &=&  \vec{v} \cdot \left( \Gamma_{\vec{P}} \vec{P} + \Gamma_{\vec{\mathbbm{P}}} \vec{\mathbbm{P}} \right)  \nonumber \\
	 				    &=& \left( \rho_{\mathrm{m}} \Gamma_{\vec{P}} + \beta_{\mathbbm{P}} \Gamma_{\vec{\mathbbm{P}}}\right) \vec{v}^2 + \left( \beta_{\mathbbm{P}} \Gamma_{\vec{P}} + \kappa_{\mathbbm{E}} \Gamma_{\vec{\mathbbm{P}}} \right) \vec{v} \cdot \vec{\mathbbm{E}} \; . 
	\end{eqnarray}
At the moment this relation appears as a condition on hydrostatic flows. When we discuss entropy production, we will show that it must also hold for non-hydrostatic flows if the ideal fluid is to not generate entropy. This relation between the $\Gamma$s can also be justified by our Drude calculation but moreover, it is simply imposing that whatever hydrodynamic theories we are looking at, they will be consistent with a stationary configuration defined by the Drude constraint Eq.~\eqref{Eq:VelocityRelation}. }

{\ Assuming $\rho_{m} \Gamma_{\vec{P}} + \beta_{\mathbbm{P}} \Gamma_{\vec{\mathbbm{P}}} \neq 0$ we can rearrange \eqref{Eq:MomentumDissipationDef} to find
	\begin{eqnarray}
	 \label{Eq:VelocityRelation2}
	 \vec{v} &=&\left( \frac{n - \kappa_{\mathbbm{E}} \Gamma_{\vec{\mathbbm{P}}} - \beta_{\mathbbm{P}} \Gamma_{\vec{P}}}{ \beta_{\mathbbm{P}} \Gamma_{\vec{\mathbbm{P}}} + \rho_{\mathrm{m}} \Gamma_{\vec{P}}} \right) \vec{\mathbbm{E}}
	 		   - \frac{n}{ \beta_{\mathbbm{P}} \Gamma_{\vec{\mathbbm{P}}} + \rho_{\mathrm{m}} \Gamma_{\vec{P}}}  \vec{\partial} \mu  + \mathcal{O}(\partial) \; ,
	\end{eqnarray}
giving us the most general hydrodynamic Drude constraint with
	\begin{subequations}
	\label{Eq:Omegaidentify}
	\begin{eqnarray}
		\label{Eq:alphaidentify}
		\Omega_{\mathbbm{E}} &=&\left( \frac{n - \kappa_{\mathbbm{E}} \Gamma_{\vec{\mathbbm{P}}} - \beta_{\mathbbm{P}} \Gamma_{\vec{P}}}{ \beta_{\mathbbm{P}} \Gamma_{\vec{\mathbbm{P}}} + \rho_{\mathrm{m}} \Gamma_{\vec{P}}} \right)  \; , \\
		\Omega_{\mu}  &=&\frac{n}{ \beta_{\mathbbm{P}} \Gamma_{\vec{\mathbbm{P}}} + \rho_{\mathrm{m}} \Gamma_{\vec{P}}} \; ,
	\end{eqnarray}
	\end{subequations}
Note that $\Omega_{\mu} \neq 0$, unless $n=0$ or $ \beta_{\mathbbm{P}} \Gamma_{\vec{\mathbbm{P}}} + \rho_{\mathrm{m}} \Gamma_{\vec{P}} \rightarrow \infty$. This shows, that charge transport away from charge neutrality requires $\Omega_{\mu} \neq 0$, thus justifying our rejection of Eq.~\eqref{Eq:VelocityNomodification} and maintaining the hydrostatic constraint $\mathbb{E}^i = \partial^i\mu$. Moreover, if we do impose \eqref{Eq:Usualconstraint}, then we recover \eqref{Eq:VelocityNomodification}, showing our result captures all cases. The Drude constraint can be re-arranged as a constraint for the relaxation terms, which might be of more use to the experimentalist, as
		\begin{eqnarray}
			\label{Eq:IndentificationGammaP}
			\left( \begin{array}{c}
				\Gamma_{\vec{\mathbbm{P}}} \\ \Gamma_{\vec{P}}
			\end{array} \right)
			&=& n \left( \begin{array}{cc}
						\beta_{\mathbbm{P}} \Omega_{\mathbbm{E}} + \kappa_{\mathbbm{E}} & \rho_{\mathrm{m}} \Omega_{\mathbbm{E}} + \beta_{\mathbbm{P}} \\ \beta_{\mathbbm{P}} \Omega_{\mu} & \rho_{\mathrm{m}} \Omega_{\mu}
					\end{array} \right)^{-1}
				\left( \begin{array}{c}
					1 \\ 1
			\end{array} \right) \; . 
		\end{eqnarray}	
This shows that our relaxation terms are completely fixed by the thermodynamics and, in particular, that both of them vanish when $n \rightarrow 0$.}

\section{Relaxed hydrodynamic equations to first order in derivatives}\label{sec:1stOrderHydro}

{\noindent We now turn to relaxation for boost agnostic fluids at order one in derivatives. In hydrodynamics without relaxation the zeroth order constitutive relations lead to first order equations of motion that are satisfied up to second order in derivatives. When we have relaxation terms, a consistent set of first order equations of motion must allow for first order pieces in the relaxation terms. Moreover, $P_{i}$ and $\mathbbm{P}_{i}$ entering in \eqref{Eq:MomentumDissipationDef} receive first order corrections; thus it is imperative to determine the forms of these corrections.}

{\ As in the previous section, we first derive the constitutive relations without momentum relaxation or imposing \eqref{Eq:VelocityRelation} via the generating functional $W_{(1)}$. In principle, $W_{(1)}$ is built out of the following independent scalar quantities
\begin{eqnarray}
	 \label{Eq:OrderOneListbeforetilde}
	 s_{(1)} &=& \left\{ \nu^{\mu} \partial_{\mu} (T,\mu,\vec{v}^2,\vec{\mathbbm{E}}^2,\vec{v} \cdot \vec{\mathbbm{E}}), \; \mathbbm{E}_{\nu} h^{\nu \mu} \partial_{\mu} (T,\mu,\vec{v}^2,\vec{\mathbbm{E}}^2,\vec{v} \cdot \vec{\mathbbm{E}}),  \;  \right. \nonumber \\
	 	   &\;& \left. \; u^{\mu} u^{\nu} \nabla_{\mu} \mathbbm{E}_{\nu}, \;  \nu^{\mu} u^{\nu} \nabla_{\mu} \mathbbm{E}_{\nu},  \;  h^{\mu \zeta} \mathbbm{E}_{\zeta} u^{\nu} \nabla_{\mu} \mathbbm{E}_{\nu}, \;  h^{\mu \nu} \nabla_{\mu} \mathbbm{E}_{\nu}, \;  \right. \nonumber \\
	 &\;& \left. \; \tau_{\rho} \Gamma_{[\mu \nu]}^{\rho} \nu^{\mu} h^{\nu \sigma} \mathbbm{E}_{\sigma} \; , \; h_{\rho \sigma} u^{\sigma} \Gamma_{[\mu \nu]}^{\rho} \nu^{\mu} h^{\nu \sigma} \mathbbm{E}_{\sigma} \right\}  \; . \nonumber
\end{eqnarray}
We have written everything in terms of covariant derivatives, in order to find the most general result, but we immediately take the flat space limit after varying. With regards to the variations, the torsion tensors are a little inconvenient to work with. As such we replace them using the following equations
	\begin{eqnarray}
		\Gamma_{[\mu \nu]}^{\rho} \tau_{\rho}
	 &=& \partial_{[\mu} \tau_{\nu]} \; , \\
			\Gamma^{\rho}_{[\mu \nu]} h_{\rho \sigma} u^{\sigma}
		&=& h_{\sigma [\nu} \nabla_{\mu]} u^{\sigma} - \partial_{[ \mu} ( h_{\nu ] \sigma} u^{\sigma} ) \; . 
	\end{eqnarray}
These express particular contractions of the torsion tensor in terms of other quantities.\footnote{In writing the list \eqref{Eq:OrderOneListbeforetilde}, it was already necessary to show that $\Gamma^{\rho}_{[\mu \nu]} h_{\rho \sigma} u^{\sigma}$ could be expressed in terms of other scalars already present in the list.} Subsequently we use the following basis of scalars
	\begin{eqnarray}
	 \label{Eq:OrderOneList}
	\tilde{s}_{(1)} &=& \left\{ \nu^{\mu} \partial_{\mu} (\mathrm{scalars}), \; \mathbbm{E}_{\nu} h^{\nu \mu} \partial_{\mu} (\mathrm{scalars}),  \; u^{\mu} u^{\nu} \nabla_{\mu} \mathbbm{E}_{\nu}, \; \nu^{\mu} u^{\nu} \nabla_{\mu} \mathbbm{E}_{\nu},  \;  h^{\mu \zeta} \mathbbm{E}_{\zeta} u^{\nu} \nabla_{\mu} \mathbbm{E}_{\nu} \; ,  \right. \nonumber \\
	 &\;& \left. \nabla_{\mu} \mathbbm{E}^{\mu}, \; \partial_{\mu} \tau_{\nu}  \nu^{[\mu} h^{\nu] \sigma} \mathbbm{E}_{\sigma} \; , \;  \partial_{ \mu} ( h_{\nu \sigma} u^{\sigma} ) \nu^{[\mu}  h^{\nu] \rho} \mathbbm{E}_{\rho} \right\} \; . 
	 \end{eqnarray}
}

{\ With the order one scalars in hand, the hydrostatic generating functional $W_{(1)}$ is given by
	\begin{eqnarray}
	\label{Eq:Orderonegenerating}
	 W_{(1)} = \int d^{d+1}x \; e \; \sum_{i} F_{i}(T,\mu, h^{\mu \nu} \mathbbm{E}_{\mu} \mathbbm{E}_{\nu}, h_{\mu \nu} u^{\mu} u^{\nu} , \mathbbm{E}_{\mu} u^{\mu}) \tilde{s}_{(1)}^{(i)} \; . 
	\end{eqnarray}
As before, varying the generating functional with respect to the clock-form, spatial metric, gauge field and field strength gives the constitutive relations \eqref{Eq:Defof1pt}. As we will only care about expressions in FSCC we can simplify our analysis by considering the FSCC limit at the level of $W_{(1)}$. In particular, because torsion vanishes in this limit the stationarity constraints that we used to eliminate torsion terms in writing \eqref{Eq:OrderOneListbeforetilde} now become relations between other variables. Namely,
	\begin{subequations}
	\label{Eq:FSCCsimpl}
	\begin{eqnarray}
		\tilde{s}_1&=&\tilde{s}_2=s_5=\tilde{s}_{13}=\tilde{s}_{14}=0 \; , \\
    		\tilde{s}_3&=&2\tilde{s}_8=2\tilde{s}_{11}=v^i\partial_i \vec{\mathbbm{E}}^2 = \bar{s}_{1} \; , \\
    		\tilde{s}_4&=&\tilde{s}_{10}=v^i\partial_i( \vec{\mathbbm{E}} \cdot \vec{v}) = \bar{s}_{2} \; , \\
    		\tilde{s}_6&=&2\tilde{s}_9= \mathbbm{E}^i\partial_i \vec{v}^2 = \bar{s}_{3} \; , \\
    		\tilde{s}_7&=& \mathbbm{E}^i\partial_i \vec{\mathbbm{E}}^2 = \bar{s}_{4} \; , \\
    		\tilde{s}_{12}&=&\partial_i \mathbbm{E}^i = \bar{s}_{5} \; . 
	\end{eqnarray}
	\end{subequations}
In addition, it is helpful to construct a basis of non-composite independent vectors in the FSCC limit. We use as a basis the following collection of vectors
	\begin{subequations}
		\label{Eq:OrderOneVectorList}
		\begin{eqnarray}
		\bar{v}_{1} &=& \partial_{i} \vec{v}^2 \; , \\
		\bar{v}_{2} &=& \partial_{i} \vec{\mathbbm{E}}^2 \; , \\
		\bar{v}_{3} &=& \partial_{i} \left( \vec{v} \cdot \vec{\mathbbm{E}} \right) \; , \\
		\bar{v}_{4} &=& v^{j} \partial_{i} \mathbbm{E}_{j} \; . 	
		\end{eqnarray}
	\end{subequations}
}

{\ Using \eqref{Eq:Defof1pt} and varying $W_{(1)}$ we obtain the constitutive relations at order one. For our purposes we shall only need to know the expressions for the spatial momentum, the polarisation, the number density  and the energy density. Henceforth, $P_{i}$, $\mathbbm{P}_{i}$ and $\varepsilon$ refer to objects with both order zero and order one terms. All other quantities defined previously take only their order zero form e.g. $n = \partial P/\partial \mu$. Given that the expressions are long and not particularly illuminating we only report these expansions and relegate them to appendix \ref{appendix:constitutive}.}

{\ We remark that in deriving the first order constitutive relations, we must deal with the issue of derivatives of the chemical potential. In \cite{Kovtun:2016lfw}, for an order zero in derivatives electric field, the derivative of the chemical potential must also be order zero. Consequently, when taking the derivative of $F_{i}$ it is possible to produce an order zero (e.g. $\partial_{\mu} F_{i} \partial_{i} \mu$) term, when we would generically expect only order one terms. Two potential interpretations for this issue were given in \cite{Kovtun:2016lfw}. We could accept that at each order in derivatives of the effective action there will be terms that contribute to the constitutive relations at one lower order. Consequently to know the constitutive relations at order $n$ we must compute the generating functional to order $n+1$. Alternatively, we can additionally assume that $\frac{\partial F_{i}}{\partial \mu} \sim \mathcal{O}(\partial)$. In \cite{Kovtun:2016lfw} this is interpreted as requiring that the effect of free charges is less important than the effect of bound charges, but arguably requires us to assume the existence of another scale in our problem.}

{\ We however can explore a new option; it is consistent for us to treat $\partial \mu \sim \mathcal{O}(\partial)$. In \cite{Kovtun:2016lfw}, $\partial^i\mu$ had to be considered as zeroth order to achieve a consistent hydrostatic equilibrium by balancing the effects of a non-zero electric fields. However, in our work we include relaxation terms, which can also balance the effect of the electric field. Therefore we can take derivatives of the chemical potential to be order one in the derivative expansion, independently of the order of the electric field. In this work we shall simply keep derivatives of the chemical potential in our expressions, trusting the reader to replace them as they please.}

\subsection{The zoology of relaxation at first order}

{\noindent To understand relaxation at first order, let us consider the momentum (non-)conservation equation written schematically as
	\begin{eqnarray}
	\label{Eq:Momentumrelaxation}
          	0 &=& \partial_{t} P_{i} + \partial_{j} T\indices{^j_i} - \left( n - \partial_{j} \mathbbm{P}^{j} \right) \mathbbm{E}_{i} + \hat{\Gamma}_{\vec{P}}^{i} + \mathcal{O}(\partial^2) \; ,
	\end{eqnarray}
Again, following the principles of effective field theory we expect that the relaxation terms must be written in terms of the effective operators up to and including order one, i.e.
	\begin{eqnarray}
		\hat{\Gamma}_{\vec{P}}^{i} &=& \left( \Gamma_{\vec{P}} + \sum_{j=1}^{5} \Gamma^{(\bar{s})}_{\vec{P},j} \bar{s}_{j} \right) P^{i} +  \left( \Gamma_{\vec{\mathbbm{P}}}  + \sum_{j=1}^{5} \Gamma^{(\bar{s})}_{\vec{\mathbbm{P}},j} \bar{s}_{j} \right) \mathbbm{P}^{i} + \sum_{j=1}^{4} \Gamma_{j}^{(\bar{v})} (\bar{v}_{j})^{i} + \mathcal{O}(\partial^2) \; . \qquad
	\end{eqnarray}
To proceed, we note that satisfying Eq.~\eqref{Eq:Momentumrelaxation} in the presence of relaxation terms, begets us to make a crucial choice; which of the stationarity constraints in Eq.~\eqref{Eq:OrderZeroConstraints} should we violate?}

{\ At order zero, we could only break the constraint \eqref{Eq:ElectricFieldBalance}, $\mathbb{E}^i = \partial^i\mu$, but at order one we have the potential to relax other hydrostaticity constraints.\footnote{One can confirm that all hydrostaticity constraints in \eqref{Eq:OrderZeroConstraints} are naively order one in derivatives.} Which is the largest set of broken constraints leading to consistent hydrodynamic equations is an open question. Here we make a much less severe modification, i.e. we continue to modify only \eqref{Eq:ElectricFieldBalance}. Then \eqref{Eq:Momentumrelaxation} evaluates to
	\begin{eqnarray}
			\label{Eq:Generichydrostaticterm}
			n\left(\mathbbm{E}_{i} - \partial_{i} \mu\right) 
		&=&  \left( \Gamma_{\vec{P}} + \sum_{j=1}^{5} \Gamma^{(\bar{s})}_{\vec{P},j} \bar{s}_{j} \right) P_{i} +  \left( \Gamma_{\vec{\mathbbm{P}}}  + \sum_{j=1}^{5} \Gamma^{(\bar{s})}_{\vec{\mathbbm{P}},j} \bar{s}_{j} \right) \mathbbm{P}_{i} \nonumber \\
		&\;& + \sum_{j=1}^{4} \Gamma_{j}^{(\bar{v})} (\bar{v}_{j})_{i} + \mathcal{O}(\partial^2) \; . 
	\end{eqnarray}
In principle, we can think of Eq.~\eqref{Eq:Generichydrostaticterm} as the definition of the broken constraint \eqref{Eq:ElectricFieldBalance} at order one in derivatives. In fact, by choosing to relax the zeroth order stationarity constraint \eqref{Eq:ElectricFieldBalance} as\footnote{Such a decomposition is conventional in the literature, allowing one to interpret the coefficients $\Gamma_{\vec{P}}$ and $\Gamma_{\vec{\mathbbm{P}}}$ as inverse relaxation times.
}
	\begin{eqnarray}
			n\left(\mathbbm{E}_{i} - \partial_{i} \mu\right) 
		&=& \Gamma_{\vec{P}} P^{i} +  \Gamma_{\vec{\mathbbm{P}}} \mathbbm{P}^{i} \; , 
	\end{eqnarray}
we had implicitly accepted that there are generically order one (and higher) corrections to the stationarity constraint. This follows from the fact that the expressions for $P_{i}$ and $\mathbbm{P}_{i}$ in terms of our operator basis receive corrections order by order (see appendix \ref{appendix:constitutive}). If we wanted to avoid first order corrections in \eqref{Eq:ElectricFieldBalance} we could choose
	\begin{eqnarray}
		\left. \Gamma_{\vec{P}} \right|_{ \mathcal{O}(\partial^{0})} &=& \Gamma_{\vec{v}} v^{i} + \Gamma_{\vec{\mathbbm{E}}} \mathbbm{E}^{i}  \; .	
	\end{eqnarray}
Doing so means we lose the interpretation of $\Gamma_{\vec{P}}$ and $\Gamma_{\vec{\mathbbm{P}}}$  as inverse relaxation times beyond lowest order, which is something we deem unphysical. It is interesting to note, however, that moving between these choices is a matter of redefining the relaxation term $\hat{\Gamma}_{\vec{P}}^{i}$ by choosing a basis for its expansion.}

{\ Moving on to the relaxed energy-conservation equation, we have at order one
	 	\begin{eqnarray}
		\label{Eq:Energyrelaxation}
		0 &=& \partial_{t} \varepsilon + \partial_{i} J^{i}_{\varepsilon} - \mathbbm{E}_{i} J^{i} 
			  + \hat{\Gamma}_{\varepsilon} \; , 
	\end{eqnarray}
which leads to
	\begin{eqnarray}
	 	n v^{i} \left( \mathbbm{E}_{i} - \partial_{i} \mu \right) &=& \hat{\Gamma}_{\varepsilon} + \mathcal{O}(\partial^2) \; . 
	\end{eqnarray}
In other words
	\begin{eqnarray}
		\label{Eq:RelaxationMatching}
		 \hat{\Gamma}_{\varepsilon} &=& v_{i} \hat{\Gamma}_{\vec{P}}^{i}  \; .  
	\end{eqnarray}
Correspondingly, this is why we chose to write our energy relaxation term without the conventional factor of the energy density. If we want to restore this energy factor, we can interpret the leading term as energy relaxation rate and write
	\begin{eqnarray}
			 \hat{\Gamma}_{\varepsilon} &=& \Gamma_{\varepsilon} \varepsilon + \sum_{j=1}^{5} \Gamma^{(\bar{s})}_{\varepsilon,j} \bar{s}_{j} + \Gamma^{(\bar{s})}_{\varepsilon,6}  v^{i} \partial_{i} \mu  + \Gamma^{(\bar{s})}_{\varepsilon,7}  \mathbbm{E}^{i} \partial_{i} \mu + \mathcal{O}(\partial^2) \; .
	\end{eqnarray}
Employing \eqref{Eq:RelaxationMatching} and \eqref{Eq:Generichydrostaticterm} we can then identify the $\Gamma^{(\bar{s})}_{\varepsilon,j}$ precisely,
	\begin{subequations}
	\label{Eq:EnergyRelaxationTimeIdentification}
	\begin{eqnarray}
		 \Gamma^{(\bar{s})}_{\varepsilon,1} &=&  \Gamma_{\vec{P}} \left( \gamma_{P,2} \vec{v}^2 + \gamma_{P,1} \vec{v} \cdot \vec{\mathbbm{E}} \right) + \Gamma_{\vec{\mathbbm{P}}} \left( \gamma_{\mathbbm{P},1} \vec{v}^2 + \gamma_{\mathbbm{P},2} \vec{v} \cdot \vec{\mathbbm{E}} \right) \nonumber \\
		 &\;& + \Gamma_{\vec{P}} \gamma_{P,17} - \Gamma_{\varepsilon} \gamma_{\varepsilon,1} \; , \\
		\Gamma^{(\bar{s})}_{\varepsilon,2} &=&  \Gamma_{\vec{P}} \left( \gamma_{P,4} \vec{v}^2 + \gamma_{P,3} \vec{v} \cdot \vec{\mathbbm{E}} \right) + \Gamma_{\vec{\mathbbm{P}}} \left( - 2 \gamma_{\mathbbm{P},1}  \vec{v} \cdot \vec{\mathbbm{E}} \right) \nonumber \\
		&\; & + \Gamma_{\vec{P}} \gamma_{P,16} - \Gamma_{\vec{\mathbbm{P}}} \gamma_{P,7} - \Gamma_{\varepsilon} \gamma_{\varepsilon,2} \; , \\
		\Gamma^{(\bar{s})}_{\varepsilon,3} &=&  \Gamma_{\vec{P}} \left( \gamma_{P,6} \vec{v}^2 + \gamma_{P,5} \vec{v} \cdot \vec{\mathbbm{E}} \right) + \Gamma_{\vec{\mathbbm{P}}} \left( \gamma_{\mathbbm{P},4} \vec{v}^2 + \gamma_{\mathbbm{P},5} \vec{v} \cdot \vec{\mathbbm{E}} \right) \nonumber \\
		&\;  &  + \frac{1}{2} \Gamma_{\vec{P}} \left( \gamma_{P,15} + \gamma_{P,16} \right)  - \Gamma_{\varepsilon} \gamma_{\varepsilon,3} \; , \\
		 \Gamma^{(\bar{s})}_{\varepsilon,4} &=&  \Gamma_{\vec{P}} \left( \gamma_{P,8} \vec{v}^2 + \gamma_{P,7} \vec{v} \cdot \vec{\mathbbm{E}} \right) + \Gamma_{\vec{\mathbbm{P}}} \left( \gamma_{\mathbbm{P},2} \vec{v}^2 \right) \nonumber \\
		&\;  &  - \Gamma_{\varepsilon} \gamma_{\varepsilon,4} \; , \\
		\Gamma^{(\bar{s})}_{\varepsilon,5} &=&  \Gamma_{\vec{P}} \left( \gamma_{P,10} \vec{v}^2 + \gamma_{P,9} \vec{v} \cdot \vec{\mathbbm{E}} \right) + \Gamma_{\vec{\mathbbm{P}}} \left( \gamma_{\mathbbm{P},7} \vec{v}^2 + \gamma_{\mathbbm{P},8} \vec{v} \cdot \vec{\mathbbm{E}} \right) \nonumber \\
		&\;  &  - \Gamma_{\varepsilon} \gamma_{\varepsilon,5} \; , \\
		\Gamma^{(\bar{s})}_{\varepsilon,6} &=&  - 2 \Gamma_{\vec{P}} \left( \gamma_{P,12} \vec{v}^2 + \gamma_{P,11} \vec{v} \cdot \vec{\mathbbm{E}} \right) - \Gamma_{\vec{\mathbbm{P}}} \left( \gamma_{\mathbbm{P},11} \vec{v}^2 + 2 \partial_{\mu} F_{3} \vec{v} \cdot \vec{\mathbbm{E}} \right) \nonumber \\
		&\;  & - \Gamma_{\vec{\mathbbm{P}}} \partial_{\mu} F_{12} - \Gamma_{\varepsilon} \gamma_{\varepsilon,6} \; , \\
		\Gamma^{(\bar{s})}_{\varepsilon,7} &=& - 2 \Gamma_{\vec{P}} \left( \gamma_{P,14} \vec{v}^2 + \gamma_{P,13} \vec{v} \cdot \vec{\mathbbm{E}} \right) - \Gamma_{\vec{\mathbbm{P}}} \left( \gamma_{\mathbbm{P},12} \vec{v}^2 + 2 \partial_{\mu} F_{7} \vec{v} \cdot \vec{\mathbbm{E}} \right) \nonumber \\
		&\;  & - \Gamma_{\vec{\mathbbm{P}}} \partial_{\vec{\mathbbm{E}}^2} F_{12} - \Gamma_{\varepsilon} \gamma_{\varepsilon,7} \; , 
	\end{eqnarray}
	\end{subequations}
where
	\begin{eqnarray}
		\Gamma_{\varepsilon} = \frac{\left( \Gamma_{\vec{P}} \rho_{\mathrm{m}} + \Gamma_{\vec{\mathbbm{P}}} \beta_{\mathbbm{P}} \right) \vec{v}^2 + \left( \Gamma_{\vec{P}} \beta_{\mathbbm{P}} + \Gamma_{\vec{\mathbbm{P}}} \kappa_{\mathbbm{E}} \right) \vec{v} \cdot \vec{\mathbbm{E}} }{ s T + \mu n + \rho_{\mathrm{m}} \vec{v}^2 + 2 \beta_{\mathbbm{P}} \vec{v} \cdot \vec{\mathbbm{E}} + \kappa_{\mathbbm{E}} \vec{\mathbbm{E}}^2 - P} \; ,
	\end{eqnarray}
where the coefficients appearing in \eqref{Eq:EnergyRelaxationTimeIdentification} are described in appendix \ref{appendix:constitutive}.}

{\ What remains to be done to complete our discussion is to examine the entropy current of our construction. Before doing so, however, we make a small observation that will be especially relevant for future studies of our systems at higher derivative orders. Consider charge relaxation
\begin{equation}
\label{Eq:SimpleRelax}
\partial_t n + \partial_i J^i = 0 \rightarrow \partial_t n + \partial_i J^i = \Gamma_{n} n~,
\end{equation}
where $\Gamma_{n}$ is the relaxation term. To both sides of this equation we can add the divergence of a vector field
\begin{equation}
\label{Eq:SimpleRelax}
\partial_t n + \partial_i \left( J^i + \Delta^{i}_{\vec{J}} \right) = \Gamma_{n} n + \partial_{i} \Delta_{\vec{J}}^{i} ~,
\end{equation}
where $\Delta_{J}^{i}$ is an example of what we name the current relaxation terms. We see that if the $\Gamma_{n}$ satisfy certain thermodynamic integrability constraints, we can use these current relaxation terms to cancel pieces of $\Gamma_{n}$. This comes at the expense of modifying our constitutive relations. In these cases the relaxation terms are in some sense fake, as they can be redefined away and the equation of motion reverts to a charge conservation equation. We do not discuss current relaxation terms in our analysis, since our $\Gamma$s will be assumed not to satisfy the requisite thermodynamic integrability constraints.}

\subsection{Entropy current conservation}

{\noindent As our conservation equations and stationarity conditions do not strictly follow from the generating functional, we should check whether the entropy current is positive definite at ideal order. Subtracting from the energy conservation equation \eqref{Eq:Energyrelaxation} the inner product of the momentum conservation equation \eqref{Eq:Momentumrelaxation} with $v^{i}$ and the charge conservation equation multiplied by $\mu$, we arrive at
    \label{Eq:EntropyProduction}
    \begin{eqnarray}
      &\;&      \left( \partial_{t} + v^{i} \partial_{i} \right) \left( \varepsilon - \vec{\mathbbm{P}} \cdot \vec{\mathbbm{E}} \right)
        - v^{i} \left( \partial_{t} + v^{j} \partial_{j} \right) P_{i}
        + \left( \varepsilon + P - \vec{\mathbbm{P}} \cdot \vec{\mathbbm{E}}
                 - v^{j} P_{j}  \right) \partial_{i} v^{i} \nonumber \\
     &\;& + v^{i} \mathbbm{P}^{j} \partial_{j} \mathbbm{E}_{i} + \mathbbm{P}^{j} \partial_{t} \mathbbm{E}^{j} \nonumber \\
     	      \label{Eq:EntropyProductionRelaxation}
       &=& \hat{\Gamma}_{\varepsilon} - v_{i} \hat{\Gamma}^{i}_{\vec{P}} + \mathcal{O}(\partial^2) \; .
    \end{eqnarray}
Using
    \begin{eqnarray}
            \left( \partial_{t} + v^{i} \partial_{i} \right) \left( \varepsilon - \vec{\mathbbm{P}} \cdot \vec{\mathbbm{E}} \right)
        &=& T \left( \partial_{t} + v^{i} \partial_{i} \right) s
            + \mu \left( \partial_{t} + v^{i} \partial_{i} \right) n \nonumber \\
        &\;& - \mathbbm{P}^{j} \left( \partial_{t} + v^{i} \partial_{i} \right) \mathbbm{E}_{j}
            + v_{j} \left( \partial_{t} + v^{i} \partial_{i} \right) P^{j}~,
     \end{eqnarray}
which follows from \eqref{Eq:InfinitesimalPressure},  this expression becomes
    \begin{eqnarray}
     \label{Eq:EntropyConservation}
    	\left( \partial_{t} + v^{i} \partial_{i} \right) s + s \partial_{i} v^{i} = \hat{\Gamma}_{\varepsilon} - v_{i} \hat{\Gamma}^{i}_{\vec{P}} + \mathcal{O}(\partial^2) \; , 
    \end{eqnarray}
where we have imposed the Bianchi identity \eqref{Eq:FlatspaceBianchi} to simplify. }{In the absence of relaxation, we know that the left hand side of \eqref{Eq:EntropyConservation} is identically satisfied and entropy is conserved. We notice that this is still the case when relaxation is included, if we use our modified hydrostatic constraint \eqref{Eq:ViolatingConservation} leading to \eqref{Eq:RelaxationMatching}. If we demand that our system only produces entropy at higher than first order in derivatives, then entropy conservation can be extended to \textit{any} solution of the hydrodynamic equations. This condition is not strictly necessary and ideal fluids can produce entropy in the presence of relaxations, as shown in \cite{Landry:2020tbh}. However the relaxations in \cite{Landry:2020tbh} are not hydrostatic and must disappear in equilibrium, while our relaxations are hydrostatic compatible, which suggest that they should not contribute to the entropy production.}

{\ It may initially seem surprising that in the presence of relaxation terms the entropy can be conserved. We remind the reader however that the second law of thermodynamics only requires entropy to increase in an \textit{isolated} system. In an open system, it may remain constant or even decrease. We can see this intuitively by examining the Gibbs-Duhem relation
	\begin{eqnarray}
	  s &=& \frac{1}{T} \left( \varepsilon - n \mu - \vec{v} \cdot \vec{P} \right) + \frac{P  - \vec{\mathbbm{E}} \cdot \vec{\mathbbm{P}}}{T} \; . 
	\end{eqnarray}
Notice that if $\vec{P}$ decreases, such as happens for the relaxation terms in \eqref{Eq:Momentumrelaxation}, we can keep the entropy constant if we simultaneously reduce the energy density (hence the necessity of an energy relaxation term). Moreover, a careful examination of how we have manipulated the equations of motion would indicate that when the derivative of the chemical potential can be treated as order one in derivatives, our entire formalism reduces to setting the source terms in energy, momentum and charge conservation equations to zero. Thus it is unsurprising that entropy is conserved.}

\subsection{Linearised stability}

{\noindent As a final check on whether our quasihydrodynamic theory is sensible, we now consider the linearised stability of our system. To do this, we must pick a hydrostatic condition. We shall make the simplest such choice, namely that
	\begin{eqnarray}
		\label{Eq:ElectricFieldBalance2}
		 n \left( \mathbbm{E}^{i} - \partial^{i} \mu \right) &=&  \Gamma_{\vec{\mathbbm{P}}} \mathbbm{P}^{i} + \Gamma_{\vec{P}} P^{i} \; .
	\end{eqnarray}
This is a minimal constraint which assumes that the Drude constraint \eqref{Eq:VelocityRelation} is exact, i.e. does not receive corrections at higher orders in derivatives.}

{\ We choose the background around which we check linear stability to be $(2+1)$-dimensional and consist of a non-zero constant chemical potential $\mu$ and electric field in the $x$-direction $\vec{\mathbbm{E}} = (\mathbbm{E}_{x},0)$. Given our constraint \eqref{Eq:ElectricFieldBalance2}, this means that the fluid also has a non-zero velocity in the $x$-direction given by
	\begin{eqnarray}
		v_{x} &= \left( \frac{n - \kappa_{\mathbbm{E}} \Gamma_{\vec{\mathbbm{P}}} - \beta_{\mathbbm{P}} \Gamma_{\vec{P}}}{ \beta_{\mathbbm{P}} \Gamma_{\vec{\mathbbm{P}}} + \rho_{\mathrm{m}} \Gamma_{\vec{P}}} \right) \mathbbm{E}_{x} \; .  	
	\end{eqnarray}
We can perturb about this background by fluctuating the temperature, chemical potential and spatial velocity. At zero wave-vector we find two zero-modes and two decaying modes. The expression for one of the decaying modes is 	\begin{eqnarray}
		\omega &=& - i \Gamma_{\mathrm{eff}.} \; , \qquad
		\Gamma_{\mathrm{eff}.} = \frac{1}{\rho_{\mathrm{m}}} \left( \beta_{\mathbbm{P}} \Gamma_{\vec{\mathbbm{P}}} + \rho_{\mathrm{m}} \Gamma_{\vec{P}} \right) \; . 
	\end{eqnarray}
The expression for the second decaying mode is significantly more complex, and in general its stability will require us to know more precisely the equation of state for the system. Nevertheless for small in amplitude electric fields the second mode has the form	
	\begin{eqnarray}
		\omega &=& - i \Gamma_{\mathrm{eff}} + \mathcal{O}(E_{x}^2) \; . 
	\end{eqnarray}
In this case our system is linearly stable on the condition $\Gamma_{\rm eff.}\geq 0 \Leftrightarrow \beta_{\mathbbm{P}} \Gamma_{\vec{\mathbbm{P}}} + \rho_{\mathrm{m}} \Gamma_{\vec{P}} \geq 0$. We have also checked that the zero modes become propagating modes at non-zero wavevector.}

{\ An interesting consequence of the stability condition is that it allows one of the relaxation rates to be negative. This allows us to tune $\Gamma_{\vec{\mathbbm{P}}}$ carefully, such that the Drude constraint \eqref{Eq:VelocityRelation} on the velocity depends solely on the total electric field $\vec{\mathbb{E}}+\vec{\mathbb{P}}-\vec{\partial}\mu$. This incorporates the backreaction of the non-zero polarisation into the problem. In particular, define
	\begin{eqnarray}
		\label{Eq:velocitytotalpolrelax}
		\Gamma_{\vec{\mathbbm{P}}} &=& - \left( n + \frac{\beta_{\mathbbm{P}}  \Gamma_{\vec{P}}}{\kappa_{\mathbbm{E}}} \right) \; . 
	\end{eqnarray}
In this case, the spatial velocity depends on the total electric field as 
	\begin{eqnarray}
		\label{Eq:velocitytotal}
		\vec{v} &=& \frac{n}{\Gamma_{\vec{P}} \left( \rho_{\mathrm{m}} - \frac{\beta_{\mathbbm{P}}^2}{\kappa_{\mathbbm{E}}} \right)}   \left( \vec{\mathbbm{E}} + \vec{\mathbbm{P}}
	 		   - \vec{\partial} \mu \right)  + \mathcal{O}(\partial)  \; ,
	\end{eqnarray}
on the condition that
	\begin{eqnarray}
		\Gamma_{\vec{P}} & \geq & \frac{n \beta_{\mathbbm{P}}}{\rho_{\mathrm{m}} - \frac{\beta_{\mathbbm{P}}^2} {\kappa_{\mathbbm{E}}}} \; , 
	\end{eqnarray}
which follows from the stability condition $\beta_{\mathbbm{P}} \Gamma_{\vec{\mathbbm{P}}} + \rho_{\mathrm{m}} \Gamma_{\vec{P}} \geq 0$. We see that this constraint reduces to the usual one $\Gamma_{\vec{P}}\geq 0$ when the compressibility of polarisation $\beta_{\mathbb{P}}$ is zero. }

{\ On physical grounds one might expect \eqref{Eq:velocitytotalpolrelax} and subsequently \eqref{Eq:velocitytotal} to hold for all systems, as it simply states that the flow velocity depends on the actual electric field experienced by the fluid particle. Yet we are not able to prove this result and do not assume it henceforth.}

\subsection{DC conductivities}\label{sec:Conductivities}

{\noindent Finally, we note that the hydrodynamic systems we have described thus far exhibit both charge and heat transport on hydrostatic solutions. In particular,  evaluating the charge current on the Drude constraint \eqref{Eq:VelocityRelation} we find 
	\begin{eqnarray}
		\label{Eq:OrderZeroDrudeJ}
		\vec{J} &=& \frac{n^2 \tau}{\rho_{\mathrm{m}}}  \vec{\mathbbm{E}} + \mathcal{O}(\partial) \; , 
	\end{eqnarray}
where
	\begin{eqnarray}
		\tau &=& \Gamma_{\vec{P}}^{-1} \left( \frac{1 - \frac{\kappa_{\mathbbm{E}} \Gamma_{\vec{\mathbbm{P}}}}{n} - \frac{\beta_{\mathbbm{P}} \Gamma_{\vec{P}}}{n}}{1 + \frac{\beta_{\mathbbm{P}} \Gamma_{\vec{\mathbbm{P}}}}{\rho_{\mathrm{m}} \Gamma_{\vec{P}}}}  \right) \; , 
	\end{eqnarray}
is the effective decay rate. This shows that our stationary states exhibit a non-trivial charge flow as well as a DC conductivity resembling the Drude result. In particular, because $\Gamma_{\vec{P}, \vec{\mathbb{P}}}$ depend on the electric field, we can define two kinds of DC conductivities: The non-linear DC conductivity, $\sigma_{\mathrm{DC}}$, is defined via 
	\begin{eqnarray}
		\label{Eq:OrderZeroDrudesigma}
		\sigma_{\mathrm{DC}} = \vec{J}/\vec{E} = \frac{n^2 \tau}{\rho_{\mathrm{m}}}  \; , 
	\end{eqnarray}
while we can also define a DC conductivity $\sigma_{\rm DC}' = \partial \vec{J}/\partial \vec{E}$. In this case we find
	\begin{eqnarray}
	\label{Eq:SigmaDifference}
		\sigma_{\mathrm{DC}}' - \sigma_{\mathrm{DC}} = \vec{\mathbbm{E}} \cdot \frac{\partial}{\partial \vec{\mathbbm{E}}} \left(  \frac{n^2 \tau}{\rho_{\mathrm{m}}} \right)
 	\end{eqnarray}
In the low electric field limit, the right hand side of \eqref{Eq:SigmaDifference} depends linearly on $\vec{v} \cdot \vec{\mathbbm{E}}$ and therefore vanishes in the limit of zero electric field, as expected. Importantly, we must remember that our results, \eqref{Eq:OrderZeroDrudeJ} and \eqref{Eq:OrderZeroDrudesigma}, depend on the ground state velocity and the applied electric field. Moreover, the heat current of our system, which can be read off from \eqref{Eq:EntropyConservation}, is given by
	\begin{eqnarray}
		\vec{Q} &=& s T \vec{v} = \frac{s T n \tau}{\rho_{\mathrm{m}}} \vec{\mathbbm{E}}~,
	\end{eqnarray}
implying our system also exhibits a non-trivial DC thermoelectric coefficient
	\begin{eqnarray}
	 \alpha_{\mathrm{DC}} &=& \frac{s T n \tau}{\rho_{\mathrm{m}}} \; . 
	\end{eqnarray}
}

{\ Before turning to the discussion, let us comment on the physical relevance of our results to potential experiments. In particular, we believe our framework should be used to describe DC transport measurements in the hydrodynamic regime of electronic materials in the presence of a constant electric field, which have reached a steady state. To understand this better, let us first remind ourselves of how the Drude form for the conductivity is derived from a naive kinetic theory calculation. Suppose one has a collection of charged, weakly interacting, slowly moving particles in the presence of a constant external electric field $\vec{\mathbbm{E}}$ that has achieved a steady state flow. Assuming that we can ignore inter-particle interactions, Newton's law tells us that the acceleration of any single charged particle is
	\begin{eqnarray}
		\label{Eq:NewtonDrude}
		m \frac{d}{dt} \langle \vec{v} \rangle = q \vec{\mathbbm{E}} - \frac{m}{\tau} \langle \vec{v} \rangle \; , 	
	\end{eqnarray}
where $\langle \vec{v} \rangle$ is the net average velocity, $m$ the particle mass and $q$ the particle charge. We have added an effective collision time $\tau$ to the system which may be treated as a phenomenological parameter describing interactions of the charges with impurities, lattices et cetera. It follows that if the system achieves a steady state flow, which can only happen when $\tau \neq \infty$, then $ d \langle \vec{v} \rangle / dt =0$ and therefore $\langle \vec{v} \rangle = q \tau / m \vec{\mathbbm{E}}$. It cannot be stressed enough that this \textit{drift velocity is a fundamental property of the material}; not the initial conditions nor the geometry of the sample - just as for our fluid discussed above. Further, assuming that the volume density of the cloud of particles is $n$, such that the spatial charge current is given by $\vec{J} = n q \langle \vec{v} \rangle$, one quickly finds that
	\begin{eqnarray}
		\label{Eq:Drudeconductivity}
		\vec{J} = \frac{n q^2 \tau}{m} \vec{\mathbbm{E}} \qquad \mathrm{such \; that} 	\qquad \sigma_{\mathrm{DC}} = \frac{n q^2 \tau}{m} \; . 
	\end{eqnarray}
}

{\ Generally when an electric current is passed through an electrical device it will heat up. If the system eventually achieves a steady state this heat must be lost by the device at the same rate that it is produced; if it does not dissipate heat the device melts. In practice this dissipation is achieved by keeping the amplitude of the electric field small, so that the produced heat is small, and we use an excellent thermal sink. An excellent thermal sink implies that we can assume the temperature of the device is the same as the temperature of the atmosphere (one finds drift velocity in data tables is specified at a given temperature). Assuming weak correlations, one can simulate such effects in a simple Drude-like model. In particular the time dependence of the average kinetic energy of the particles is given by
	\begin{eqnarray}
		\frac{m}{2} \frac{d}{dt} \langle \vec{v} \rangle^2 = e \langle \vec{v} \rangle \cdot \vec{\mathbbm{E}} - \frac{m}{\tau} \langle \vec{v} \rangle^2 \; . 
	\end{eqnarray}
Imposing the steady flow constraint of \eqref{Eq:NewtonDrude} one finds that the right hand side of this equation is also zero. The second term in the above expression is an example energy relaxation term similar to our $\Gamma_\epsilon$ in the hydrodynamic picture.}

\section{Discussion}\label{sec:Discussion}

{\noindent In this paper we have demonstrated that it is possible to consistently modify the hydrostaticity conditions using relaxation terms up to and including order one in derivatives.  We have shown that the relaxation terms are not arbitrary, but must satisfy particular constraints; more specifically, our relaxation terms are completely fixed by the thermodynamics (see \eqref{Eq:IdentificationofGammaE} and \eqref{Eq:IndentificationGammaP}) and they vanish when the charge density $n$ goes to zero. This leads to a ground state of our fluid with a constant flow of charge and heat. These flows result in our fluid also exhibiting both DC electric charge and thermo-electric conductivities that depend non-linearly on the electric field, which we calculated explicitly.}

{\ In our work we have treated relaxation terms consistently by splitting them into two kinds - relaxation terms that can be expanded in the basis of stationary tensor structures (considered in this paper) and relaxation terms that can be expanded in terms of tensor structures that vanish at stationarity (for future work). This gives a precise meaning to the idea of relaxation term being of a particular order in derivatives; which can be ascertained by examining the order of the relevant tensor structure.}

{\ As regards the future perspectives, a fundamental question is to determine how our analysis is modified when employing the order one constitutive relations in their respective equations of motion. A consistent analysis will require us to obtain second order relaxation terms. Specifically, it would be interesting to determine how the DC conductivities are modified by these additional terms as this is also the order where tensor structures that vanish at stationarity, including the type associated with the incoherent conductivity, can be introduced. A straightforward follow up of our work would be to include in the analysis such dissipative terms, and analyse their implications at the level of thermo-electric transport.}

{\ Finally, it would be important to study holographic models which realise the hydrodynamic picture described in this paper. With a holographic realisation to hand, it would be possible to analyze further the validity of our hydrodynamic description and to understand the transition from the hydrodynamic to the non-hydrodynamic regime in the new class of fluids which we have described. Holographic models with a background electric field have been considered e.g. in \cite{Withers:2016lft}. It would be interesting to generalise this analysis in order to include relaxation terms of the kind described in this work.}

\section*{Acknowledgements}

{\noindent We would like to acknowledge discussions with Blaise Gout\'{e}raux, Ashish Shukla, Benjamin Withers and Vaios Ziogas on the issues presented in this work. A.A. and I.M. have been partially supported by the “Curiosity Driven Grant 2020” of the University of Genoa and the INFN Scientific Initiative SFT: “Statistical Field Theory, Low-Dimensional Systems, Integrable Models and Applications”. This project has also received funding from the European Union’s Horizon 2020 research and innovation programme under the Marie Sklodowska-Curie grant agreement No. 101030915. }

\bibliographystyle{JHEP}
\bibliography{refs}

\appendix

\section{Conservation equations and hydrostaticity}\label{appendix:Conserved}

{\noindent In this appendix, we show that the equations of motion of the fluid are satisfied identically when we use the stationarity constraints \eqref{Eq:OrderZeroConstraints} and \eqref{Eq:ElectricFieldBalance} in the FSCC limit. First, recall that requiring diffeomorphism and gauge invariance of the generating functional $W$, yields the following (non-)conservation equations
	\begin{subequations}
	\begin{eqnarray}
	   \label{Eq:SEMconscurved'}
	   e^{-1} \partial_{\mu} \left( e T\indices{^\mu_\rho} \right) + T^{\mu} \partial_{\rho} \tau_{\mu} - \frac{1}{2} T^{\mu \nu} \partial_{\rho} h_{\mu \nu} &=& F_{\rho \mu} J^{\mu} \; , \\
	   \label{Eq:Chargeconscurved'}
	   e^{-1} \partial_{\mu} \left( e J^{\mu} \right) &=& 0 \; ,
	\end{eqnarray}
where we have employed \eqref{Eq:Defof1pt} and \eqref{Eq:SEMtensor}. In addition we have the following conservation equation for the polarisation
	\begin{eqnarray}
		   e^{-1} \partial_{\mu} \partial_\nu\left(2e\nu^{[\mu}\mathbbm{P}^{\nu]}\right) &=& 0 \; .
	\end{eqnarray}
	\end{subequations}
}

{\ Turning to the FSCC limit the electric charge conservation equation becomes
	\begin{eqnarray}
		\partial_{t} n + \partial_{i} J^{i} = 0 \; . 
	\end{eqnarray}
Employing the constitutive relations of \eqref{Eq:ConstitutiveRelations0} we find
    \begin{eqnarray}
       \label{Eq:Chargeconsflat0}
        \left( \partial_{t} n + v^{i} \partial_{i} n \right) + n \partial_{i} v^{i} &=& 0 \; .  
    \end{eqnarray}
Setting the first term (in brackets) to zero follows from $\mathcal{L}_{u}(n)=0$. The remaining term vanishes if the expansion of the fluid, $\partial_{i} v^{i}$, is zero. This is indeed one of the conditions in \eqref{Eq:OrderZeroConstraints}.}

{\  For SEM tensor conservation, \eqref{Eq:SEMconscurved}, the time component in flat space takes the form
    \begin{eqnarray}
          \partial_{t} \varepsilon + \partial_{i} J^{i}_{\varepsilon} - \mathbbm{E}_{i} J^{i}
         = 0 \; , 
    \end{eqnarray}
where $J^{i}_{\varepsilon} = T\indices{^{i}_{t}}$ is the energy current. Substituting in the constitutive relations \eqref{Eq:ConstitutiveRelations0} we obtain
    \begin{eqnarray}
    	\label{Eq:Energyconsflat0}
            0
        &=& \left( \partial_{t} + v^{i} \partial_{i} \right) \left( \varepsilon -\vec{\mathbbm{P}} \cdot \vec{\mathbbm{E}} \right) + \frac{\kappa_{\mathbbm{E}}}{2} \left( \partial_{t} + v^{i} \partial_{i} \right) \mathbbm{E}^2 + \beta_{\mathbbm{P}}  \left( \partial_{t} + v^{i} \partial_{i} \right) \left( \vec{\mathbbm{E}} \cdot \vec{v} \right)  \nonumber \\
        &\;& + \left( \varepsilon + P - \vec{\mathbbm{P}} \cdot \vec{\mathbbm{E}} \right) \partial_{i} v^{i} + \rho_{\mathrm{m}} v^{i} v^{j} \partial_{i} v_{j}   - \beta_{\mathbbm{P}} \mathbbm{E}_{i} \partial_{t} v^{i} \nonumber \\
        &\;&  - v^{i} \left( n \left( \mathbbm{E}_{i} - \partial_{i} \mu \right) - s \partial_{i} T  \right) \; , 
    \end{eqnarray}
In the hydrostatic limit the first three terms vanish on account of requiring conservation of scalar quantities; then the next four terms vanish on account of conditions in \eqref{Eq:OrderZeroConstraints}.}

{\ Finally, we come to momentum conservation given by the spatial part of \eqref{Eq:SEMconscurved}. In flat space we find
    \begin{eqnarray}
            \partial_{t} P_{i} + \partial_{j} T\indices{^{j}_{i}}  - \left( n - \partial_{j} \mathbbm{P}^{j} \right) \mathbbm{E}_{i} = 0 \; .
    \end{eqnarray}
Once again, expanding terms by substituting in the constitutive relations we find
    \begin{eqnarray}
       \label{Eq:Momconsflat0}
     0
        &=& v_{i} \left( \partial_{t} + v^{j} \partial_{j} \right) \rho_{\mathrm{m}} + \mathbbm{E}_{i} \left( \partial_{t} + v^{j} \partial_{j} \right) \beta_{\mathbbm{P}}  + \rho_{\mathrm{m}} v_{i}  \partial_{j} v^{j} 
              \nonumber \\
        &\;& - n \left( \mathbbm{E}_{i} - \partial_{i} \mu \right) + s \partial_{i} T + \rho_{\mathrm{m}} \left( \partial_{t} v_{i} + v^{j} \left(  \partial_{i} v_{j} + \partial_{j} v_{i} \right) \right) \nonumber \\
        &\;& + \left( \kappa_{\mathbbm{E}} \mathbbm{E}^{j} + \beta_{\mathbbm{P}}v^{j} \right) \left( \partial_{i} \mathbbm{E}_{j} - \partial_{j} \mathbbm{E}_{i} \right) \nonumber \\
        &\;& + \beta_{\mathbbm{P}}  \left( \partial_{t} \mathbbm{E}_{i} + v^{j} \partial_{j} \mathbbm{E}_{i} + \mathbbm{E}_{j} \partial_{i} v^{j} \right) \; . \qquad
    \end{eqnarray}
Term by term we again find that the equation of motion is identically satisfied by the hydrostatic constraints \eqref{Eq:OrderZeroConstraints} completing our demonstration.}

\section{Transport coefficients in FSCC}
\label{appendix:constitutive}

\subsection{Energy density}
\label{appendix:energy}

{\noindent The energy density $\varepsilon$ up to and including order one in derivatives has the form
	\begin{eqnarray}
			\varepsilon
		&=& - P + sT+\mu n+  \rho_{\mathrm{m}} \vec{v}^2 + 2 \beta_{\mathbbm{P}} \vec{v} \cdot \vec{\mathbbm{E}} + \kappa_{\mathbbm{E}} \vec{\mathbbm{E}}^2	\nonumber \\
		&\;& + \gamma_{\varepsilon,1} v^i \partial_i \vec{\mathbbm{E}}^2 + \gamma_{\varepsilon,2} v^i\partial_i (\vec{\mathbbm{E}} \cdot \vec{v})+\gamma_{\varepsilon,3} \mathbbm{E}^j \partial_j \vec{v}^2 + \gamma_{\varepsilon,4} \mathbbm{E}^i \partial_i \vec{\mathbbm{E}}^2+\gamma_{\varepsilon,5} \partial_i \vec{\mathbbm{E}}^i\nonumber\\
    &\;& -\gamma_{\varepsilon,6} v^i\partial_i\mu- \gamma_{\varepsilon,7} \mathbbm{E}^i\partial_i\mu \; , 
    \end{eqnarray}
where
\begin{eqnarray}
          \gamma_{\varepsilon,1}  
    &=&\mu\frac{\partial F_3}{\partial\mu}+T\left(\frac{\partial F_3}{\partial T}-\frac{\partial F_1}{\partial \vec{\mathbbm{E}}^2}\right)+2\left(\frac{\partial F_3}{\partial \vec{v}^2}-\frac{\partial F_2}{\partial \vec{\mathbbm{E}}^2}\right)\vec{v}^2+\nonumber\\
    &\;& +\left(\frac{\partial F_3}{\partial(\vec{\mathbbm{E}} \cdot \vec{v})}-\frac{\partial F_4}{\partial \vec{\mathbbm{E}}^2}\right)(\vec{\mathbbm{E}} \cdot \vec{v})-\frac{(\vec{\mathbbm{E}} \cdot \vec{v})}{2}\frac{\partial F_{14}}{\partial \vec{\mathbbm{E}}^2}+\left(\frac{\partial F_3}{\partial(\vec{\mathbbm{E}} \cdot \vec{v})}-\frac{\partial F_4}{\partial \vec{\mathbbm{E}}^2}\right)(\vec{\mathbbm{E}} \cdot \vec{v})+\nonumber\\
    &\;&-\frac{\partial F_{10}}{\partial \vec{\mathbbm{E}}^2}(\vec{\mathbbm{E}} \cdot \vec{v})+\frac{\mu}{2}\frac{\partial F_8}{\partial\mu}+\frac{T}{2}\left(\frac{\partial F_8}{\partial T}-\frac{\partial F_5}{\partial(\vec{\mathbbm{E}} \cdot \vec{v})}\right)+\left(\frac{\partial F_8}{\partial \vec{v}^2}-\frac{\partial F_6}{\partial(\vec{\mathbbm{E}} \cdot \vec{v})}\right)\vec{v}^2+\nonumber\\
    &\;&-\left(\frac{\partial F_7}{\partial(\vec{\mathbbm{E}} \cdot \vec{v})}-\frac{\partial F_8}{\partial \vec{\mathbbm{E}}^2}\right)\vec{\mathbbm{E}}^2-\frac{(\vec{\mathbbm{E}} \cdot \vec{v})}{2}\frac{\partial F_{11}}{\partial(\vec{\mathbbm{E}} \cdot \vec{v})}+\frac{1}{4}\frac{\partial F_{13}}{\partial(\vec{\mathbbm{E}} \cdot \vec{v})}+\nonumber\\
    &\;&+\frac{F_{11}}{2}+\frac{\mu}{2}\frac{\partial F_{11}}{\partial\mu}+\frac{T}{2}\frac{\partial F_{11}}{\partial T}+\frac{\partial F_{11}}{\partial \vec{v}^2}\vec{v}^2+\frac{(\vec{\mathbbm{E}} \cdot \vec{v})}{2}\frac{\partial F_{11}}{\partial(\vec{\mathbbm{E}} \cdot \vec{v})}+\nonumber\\
    &\;&+\frac{(\vec{\mathbbm{E}} \cdot \vec{v})}{2}\frac{\partial F_{11}}{\partial(\vec{\mathbbm{E}} \cdot \vec{v})}+\frac{\partial F_{11}}{\partial \vec{\mathbbm{E}}^2}\vec{\mathbbm{E}}^2\\
          \gamma_{\varepsilon,2} 
    &=&\mu\frac{\partial F_4}{\partial\mu}+T\left(\frac{\partial F_4}{\partial T}-\frac{\partial F_1}{\partial(\vec{\mathbbm{E}} \cdot \vec{v})}\right)+2\left(\frac{\partial F_4}{\partial \vec{v}^2}-\frac{\partial F_2}{\partial(\vec{\mathbbm{E}} \cdot \vec{v})}\right)\vec{v}^2-\frac{(\vec{\mathbbm{E}} \cdot \vec{v})}{2}\frac{\partial F_{14}}{\partial(\vec{\mathbbm{E}} \cdot \vec{v})}+\nonumber\\
    &\;& -2\left(\frac{\partial F_3}{\partial(\vec{\mathbbm{E}} \cdot \vec{v})}-\frac{\partial F_4}{\partial \vec{\mathbbm{E}}^2}\right)\vec{\mathbbm{E}}^2-\frac{\partial F_{10}}{\partial(\vec{\mathbbm{E}} \cdot \vec{v})}(\vec{\mathbbm{E}} \cdot \vec{v})+F_{10}+\mu\frac{\partial F_{10}}{\partial\mu}+T\frac{\partial F_{10}}{\partial T}\nonumber\\
    &\;&+2\frac{\partial F_{10}}{\partial \vec{v}^2}\vec{v}^2+\frac{\partial F_{10}}{\partial(\vec{\mathbbm{E}} \cdot \vec{v})}(\vec{\mathbbm{E}} \cdot \vec{v})+\frac{\partial F_{10}}{\partial(\vec{\mathbbm{E}} \cdot \vec{v})}(\vec{\mathbbm{E}} \cdot \vec{v})+2\frac{\partial F_{10}}{\partial \vec{\mathbbm{E}}^2}\vec{\mathbbm{E}}^2 -\frac{1}{2}F_{14} \\
    	   \gamma_{\varepsilon,3}
    &=&\mu\frac{\partial F_6}{\partial\mu}+T\left(\frac{\partial F_6}{\partial T}-\frac{\partial F_5}{\partial \vec{v}^2}\right)+\left(\frac{\partial F_6}{\partial(\vec{\mathbbm{E}} \cdot \vec{v})}-\frac{\partial F_8}{\partial \vec{v}^2}\right)(\vec{\mathbbm{E}} \cdot \vec{v})+\nonumber\\
    &\;&+\left(\frac{\partial F_6}{\partial(\vec{\mathbbm{E}} \cdot \vec{v})}-\frac{\partial F_8}{\partial \vec{v}^2}\right)(\vec{\mathbbm{E}} \cdot \vec{v})+2\left(\frac{\partial F_6}{\partial \vec{\mathbbm{E}}^2}-\frac{\partial F_7}{\partial \vec{v}^2}\right)\vec{\mathbbm{E}}^2-\frac{\partial F_{11}}{\partial \vec{v}^2}(\vec{\mathbbm{E}} \cdot \vec{v})+\frac{1}{2}\frac{\partial F_{13}}{\partial \vec{v}^2}+\nonumber\\
    &\;&+F_9+\frac{\mu}{2}\frac{\partial F_9}{\partial\mu}+\frac{T}{2}\frac{\partial F_9}{\partial T}+\frac{\partial F_9}{\partial \vec{v}^2}\vec{v}^2+\frac{\partial F_9}{\partial(\vec{\mathbbm{E}} \cdot \vec{v})}\frac{(\vec{\mathbbm{E}} \cdot \vec{v})}{2}-\frac{F_9}{2}+\nonumber\\
    &\;&+\frac{(\vec{\mathbbm{E}} \cdot \vec{v})}{2}\frac{\partial F_9}{\partial(\vec{\mathbbm{E}} \cdot \vec{v})}+\frac{\partial F_9}{\partial \vec{\mathbbm{E}}^2}\vec{\mathbbm{E}}^2 +\frac{1}{2}F_9 \\
         \gamma_{\varepsilon,4} 
    &=&\mu\frac{\partial F_7}{\partial\mu}+T\left(\frac{\partial F_7}{\partial T}-\frac{\partial F_5}{\partial \vec{\mathbbm{E}}^2}\right)+2\left(\frac{\partial F_7}{\partial \vec{v}^2}-\frac{\partial F_6}{\partial \vec{\mathbbm{E}}^2}\right)\vec{v}^2+\left(\frac{\partial F_7}{\partial(\vec{\mathbbm{E}} \cdot \vec{v})}-\frac{\partial F_8}{\partial \vec{\mathbbm{E}}^2}\right)(\vec{\mathbbm{E}} \cdot \vec{v})+\nonumber\\
    &\;&+\left(\frac{\partial F_7}{\partial(\vec{\mathbbm{E}} \cdot \vec{v})}-\frac{\partial F_8}{\partial \vec{\mathbbm{E}}^2}\right)(\vec{\mathbbm{E}} \cdot \vec{v})-\frac{\partial F_{11}}{\partial \vec{\mathbbm{E}}^2}(\vec{\mathbbm{E}} \cdot \vec{v})+\frac{1}{2}\frac{\partial F_{13}}{\partial \vec{\mathbbm{E}}^2}\\
          \gamma_{\varepsilon,5} 
    &=&\mu\frac{\partial F_{12}}{\partial\mu}+T\frac{\partial F_{12}}{\partial T}+2\frac{\partial F_{12}}{\partial \vec{v}^2}\vec{v}^2+\frac{\partial F_{12}}{\partial(\vec{\mathbbm{E}} \cdot \vec{v})}(\vec{\mathbbm{E}} \cdot \vec{v})+\frac{\partial F_{12}}{\partial(\vec{\mathbbm{E}} \cdot \vec{v})}(\vec{\mathbbm{E}} \cdot \vec{v})+2\frac{\partial F_{12}}{\partial \vec{\mathbbm{E}}^2}\vec{\mathbbm{E}}^2+\nonumber\\
    &\;& -2\vec{\mathbbm{E}}^2F_7-(\vec{\mathbbm{E}} \cdot \vec{v})F_8-TF_5-2\vec{v}^2F_6-(\vec{\mathbbm{E}} \cdot \vec{v})F_8-F_{11}(\vec{\mathbbm{E}} \cdot \vec{v})+\frac{F_{13}}{2}\\
    	   \gamma_{\varepsilon,6} 
    &=&T\frac{\partial F_1}{\partial\mu}+2\vec{v}^2\frac{\partial F_2}{\partial\mu}+(\vec{\mathbbm{E}} \cdot \vec{v})\frac{\partial F_4}{\partial\mu}+\frac{(\vec{\mathbbm{E}} \cdot \vec{v})}{2}\frac{\partial F_{14}}{\partial\mu}+2\frac{\partial F_3}{\partial\mu}\vec{\mathbbm{E}}^2+(\vec{\mathbbm{E}} \cdot \vec{v})\frac{\partial F_4}{\partial\mu}\nonumber\\
    &\;&+\frac{\partial F_{10}}{\partial\mu}(\vec{\mathbbm{E}} \cdot \vec{v})\\
    	  \gamma_{\varepsilon,7} 
    &=&T\frac{\partial F_5}{\partial\mu}+2\vec{v}^2\frac{\partial F_6}{\partial\mu}+\frac{\partial F_8}{\partial\mu}(\vec{\mathbbm{E}} \cdot \vec{v})+2\vec{\mathbbm{E}}^2\frac{\partial F_7}{\partial\mu}+\frac{\partial F_8}{\partial\mu}(\vec{\mathbbm{E}} \cdot \vec{v})+\frac{\partial F_{11}}{\partial\mu}-\frac{1}{2}\frac{\partial F_{13}}{\partial\mu} \; .
\end{eqnarray}
}

\subsection{Electric charge density}
\label{appendix:number}

{\noindent The order one in derivatives correction to the charge density, $\delta n$, has the form
	\begin{eqnarray}
		\delta n
		&=&		\left[ - \frac{\partial F_{3}}{\partial \mu} \partial_{t} \vec{\mathbbm{E}}^2
	 	- \frac{\partial F_{4}}{\partial \mu} \partial_{t} (\vec{\mathbbm{E}} \cdot \vec{v})^2 +  \frac{\partial F_{6}}{\partial \mu} \mathbbm{E}^{j} \partial_{j} \vec{v}^2  
		+  \frac{\partial F_{7}}{\partial \mu}  \mathbbm{E}^{j} \partial_{j} \vec{\mathbbm{E}}^2 \right. \nonumber \\
	 &\;& \left.  \hphantom{\frac{\partial P}{\partial \mu} + 	\left[ \right.} +  \frac{\partial F_{8}}{\partial \mu} \mathbbm{E}^{j} \partial_{j} (\vec{\mathbbm{E}} \cdot \vec{v})^2  +  \frac{\partial F_{9}}{\partial \mu} \left( v^{i} \partial_{t} \mathbbm{E}_{i} + v^{i} v^{j} \partial_{i} \mathbbm{E}_{j} \right) \right. \nonumber \\
	 &\;& \left. \hphantom{\frac{\partial P}{\partial \mu} + 	\left[ \right.}  - \frac{\partial F_{10}}{\partial \mu} v^{j} \partial_{t} \mathbbm{E}_{j} +  \frac{\partial F_{11}}{\partial \mu} v^{i} \mathbbm{E}^{j} \partial_{j} \mathbbm{E}_{i} +  \frac{\partial F_{12}}{\partial \mu} \partial_{j} \mathbbm{E}^{j} \right] + \mathcal{O}(\partial^2) \;  .
	\end{eqnarray}
}

\subsection{Spatial momentum density}
\label{appendix:momentum}

{\noindent The spatial momentum density $P_{i}$ up to and including first order in derivatives has the form
	\begin{eqnarray}
		P^{i} &=& \rho_{\mathrm{m}} v^{i} + \beta_{\mathbbm{P}} \mathbbm{E}^{i} \nonumber \\
    &\;& + \left( \gamma_{P,2} v^j\partial_j\vec{\mathbbm{E}}^2 + \gamma_{P,4} v^j\partial_j(\vec{\mathbbm{E}} \cdot \vec{v}) +  \gamma_{P,6} \mathbbm{E}^j\partial_j\vec{v}^2 + \gamma_{P,8}  \mathbbm{E}^j\partial_j\vec{\mathbbm{E}}^2 + \gamma_{P,10} \partial_j \mathbbm{E}^j  \right.  \nonumber\\
    &\;& \left. \hphantom{+ \left(  \right.} - 2 \gamma_{P,12} v^j\partial_j\mu - 2 \gamma_{P,14} \mathbbm{E}^j\partial_j\mu  \right) v^{i} \nonumber\\
    &\;& + \left( \gamma_{P,1} v^j\partial_j\vec{\mathbbm{E}}^2 + \gamma_{P,3} v^j\partial_j(\vec{\mathbbm{E}} \cdot \vec{v}) + \gamma_{P,5} \mathbbm{E}^j\partial_j\vec{v}^2 + \gamma_{P,7} \mathbbm{E}^j\partial_j\vec{\mathbbm{E}}^2 + \gamma_{P,9} \partial_j \mathbbm{E}^j \right. \nonumber\\
    &\;&  \left. \hphantom{+ \left(  \right.} -2 \gamma_{P,11} v^j\partial_j\mu -2 \gamma_{P,13} \mathbbm{E}^j\partial_j\mu  \right) \mathbbm{E}^i \nonumber\\
    &\;& +  \gamma_{P,15} \mathbbm{E}^j\partial_jv^i+  \gamma_{P,16} v^j\partial_j \mathbbm{E}^i+  \gamma_{P,17} \partial^i\vec{\mathbbm{E}}^2+F_2\partial^i\vec{v}^2 \; ,
    \end{eqnarray}
where
\begin{eqnarray}
    \gamma_{P,1}&=&\frac{\partial F_3}{\partial(\vec{\mathbbm{E}} \cdot \vec{v})}-\frac{\partial F_4}{\partial \vec{\mathbbm{E}}^2}+\frac{1}{2}\frac{\partial F_9}{\partial \vec{\mathbbm{E}}^2}-\frac{\partial F_{10}}{\partial \vec{\mathbbm{E}}^2}+\frac{1}{2}\frac{\partial F_{10}}{\partial \vec{\mathbbm{E}}^2}+\frac{1}{2}\frac{\partial F_{11}}{\partial(\vec{\mathbbm{E}} \cdot \vec{v})} -\frac{1}{2}\frac{\partial F_{14}}{\partial \vec{\mathbbm{E}}^2} \\
    \gamma_{P,2}&=&2\frac{\partial F_3}{\partial \vec{v}^2}-2\frac{\partial F_2}{\partial \vec{\mathbbm{E}}^2}+\frac{\partial F_8}{\partial \vec{v}^2}-\frac{\partial F_6}{\partial(\vec{\mathbbm{E}} \cdot \vec{v})}-\frac{1}{4}\frac{\partial F_{14}}{\partial(\vec{\mathbbm{E}} \cdot \vec{v})}-\frac{1}{2}\frac{\partial F_9}{\partial(\vec{\mathbbm{E}} \cdot \vec{v})}+\frac{1}{4}\frac{\partial F_9}{\partial(\vec{\mathbbm{E}} \cdot \vec{v})}+\nonumber\\
    &\;& +\frac{1}{4}\frac{\partial F_{10}}{\partial(\vec{\mathbbm{E}} \cdot \vec{v})}+\frac{\partial F_{11}}{\partial \vec{v}^2}\\
    \gamma_{P,3}&=&\frac{1}{2}\frac{\partial F_9}{\partial(\vec{\mathbbm{E}} \cdot \vec{v})}-\frac{\partial F_{10}}{\partial(\vec{\mathbbm{E}} \cdot \vec{v})}+\frac{1}{2}\frac{\partial F_{10}}{\partial(\vec{\mathbbm{E}} \cdot \vec{v})}+\frac{\partial F_{10}}{\partial(\vec{\mathbbm{E}} \cdot \vec{v})} -\frac{1}{2}\frac{\partial F_{14}}{\partial(\vec{\mathbbm{E}} \cdot \vec{v})} \\
    \gamma_{P,4}&=&2\frac{\partial F_4}{\partial \vec{v}^2}-2\frac{\partial F_2}{\partial(\vec{\mathbbm{E}} \cdot \vec{v})}+2\frac{\partial F_{10}}{\partial \vec{v}^2}\\
    \gamma_{P,5}&=&\frac{\partial F_6}{\partial(\vec{\mathbbm{E}} \cdot \vec{v})}-\frac{\partial F_8}{\partial \vec{v}^2}+\frac{1}{2}\frac{\partial F_9}{\partial(\vec{\mathbbm{E}} \cdot \vec{v})}\\
    \gamma_{P,6}&=&-\frac{1}{2}\frac{\partial F_{14}}{\partial \vec{v}^2}-\frac{\partial F_9}{\partial \vec{v}^2}+\frac{1}{2}\frac{\partial F_9}{\partial \vec{v}^2}+\frac{1}{2}\frac{\partial F_{10}}{\partial \vec{v}^2}+\frac{\partial F_9}{\partial \vec{v}^2}\\
    \gamma_{P,7}&=&\frac{\partial F_7}{\partial(\vec{\mathbbm{E}} \cdot \vec{v})}-\frac{\partial F_8}{\partial \vec{\mathbbm{E}}^2}\\
    \gamma_{P,8}&=&2\frac{\partial F_7}{\partial \vec{v}^2}-2\frac{\partial F_6}{\partial \vec{\mathbbm{E}}^2}-\frac{1}{2}\frac{\partial F_{14}}{\partial \vec{\mathbbm{E}}^2}-\frac{\partial F_9}{\partial \vec{\mathbbm{E}}^2}+\frac{1}{2}\frac{\partial F_9}{\partial \vec{\mathbbm{E}}^2}+\frac{1}{2}\frac{\partial F_{10}}{\partial \vec{\mathbbm{E}}^2}\\
    \gamma_{P,9}&=&\frac{\partial F_{12}}{\partial(\vec{\mathbbm{E}} \cdot \vec{v})}-F_8\\
    \gamma_{P,10}&=&2\frac{\partial F_{12}}{\partial \vec{v}^2}-F_9+\frac{1}{2}F_9+\frac{1}{2}F_{10}-2F_6-\frac{1}{2}F_{14}\\
   \gamma_{P,11}&=&\frac{1}{2}\frac{\partial F_4}{\partial\mu}-\frac{1}{4}\frac{\partial F_9}{\partial\mu}+\frac{1}{2}\frac{\partial F_{10}}{\partial\mu}-\frac{1}{4}\frac{\partial F_{10}}{\partial\mu}\\
    \gamma_{P,12}&=&\frac{\partial F_2}{\partial\mu}\\
     \gamma_{P,13}&=&\frac{1}{2}\frac{\partial F_8}{\partial\mu}\\
     \gamma_{P,14} &=&\frac{\partial F_6}{\partial\mu}+\frac{1}{4}\frac{\partial F_{14}}{\partial\mu}+\frac{1}{2}\frac{\partial F_9}{\partial\mu}-\frac{1}{4}\frac{\partial F_9}{\partial\mu}-\frac{1}{4}\frac{\partial F_{10}}{\partial\mu} \\
    \gamma_{P,15} &=& \frac{1}{2}F_9-\frac{1}{2}F_{14}-\frac{1}{2}F_{10}-F_4\\
    \gamma_{P,16} &=&\frac{1}{2}F_9-\frac{1}{2}F_{14}+\frac{3}{2}F_{10}+F_4\\
    \gamma_{P,17} &=& F_3+\frac{1}{2}F_{11} \; . 
\end{eqnarray}
}

\subsection{Polarisation}
\label{appendix:polarisation}

{\noindent The polarisation vector $\mathbbm{P}_{i}$ up to and including order one in derivatives has the form
	\begin{eqnarray}
			\label{Eq:Polarisationvectoratorderone}
			\mathbbm{P}^{i}
		&=& \beta_{\mathbbm{P}} v^{i} + \kappa_{\mathbbm{E}} \mathbbm{E}^{i} \nonumber \\
		&\;& + \left(\gamma_{\mathbbm{P},1} v^j\partial_j \vec{\mathbbm{E}}^2 + \gamma_{\mathbbm{P},4} \mathbbm{E}^j\partial_j \vec{v}^2 -\gamma_{\mathbbm{P},2} \mathbbm{E}^j\partial_j \mathbbm{E}^2 + \gamma_{\mathbbm{P},7} \partial_j \mathbbm{E}^j -\gamma_{\mathbbm{P},{11}} v^j\partial_j\mu \right. \nonumber \\
		&\;&  \left. \hphantom{+\left( \right.} - \gamma_{\mathbbm{P},12} \mathbbm{E}^j\partial_j\mu\right) v^i \nonumber\\
    &\;& +\left(\gamma_{\mathbbm{P},2} v^j\partial_j \vec{\mathbbm{E}}^2-2\gamma_{\mathbbm{P},1} v^j\partial_j( \vec{\mathbbm{E}}\cdot \vec{v}) +\gamma_{\mathbbm{P},5} \mathbbm{E}^j\partial_j \vec{v}^2+\gamma_{\mathbbm{P},8} \partial_j \mathbbm{E}^j - 2\frac{\partial F_3}{\partial\mu}v^j\partial_j\mu \nonumber \right.\\
    &\;& \left. \hphantom{+\left( \right.} -2\frac{\partial F_7}{\partial\mu} \mathbbm{E}^j\partial_j\mu\right) \mathbbm{E}^i  \nonumber \\
    &\;& + \gamma_{\mathbbm{P},9} \partial^i \vec{v}^2 - \gamma_{\mathbbm{P},7} \partial^i( \vec{\mathbbm{E}} \cdot \vec{v} )-\frac{\partial F_{12}}{\partial\mu}\partial^i\mu-\frac{\partial F_{12}}{\partial \vec{\mathbbm{E}}^2}\partial^i \vec{\mathbbm{E}}^2  \; ,
	\end{eqnarray}
where
\begin{eqnarray}
    \gamma_{\mathbbm{P},1}&=&\frac{\partial F_3}{\partial(\vec{\mathbbm{E}} \cdot \vec{v})}-\frac{\partial F_4}{\partial \vec{\mathbbm{E}}^2}-\frac{\partial F_{10}}{\partial \vec{\mathbbm{E}}^2}\\
    \gamma_{\mathbbm{P},2}&=&\frac{\partial F_8}{\partial \vec{\mathbbm{E}}^2}-\frac{\partial F_7}{\partial (\vec{\mathbbm{E}} \cdot \vec{v})}+\frac{\partial F_{11}}{\partial \vec{\mathbbm{E}}^2}\\
    \gamma_{\mathbbm{P},3}&=&2\left(\frac{\partial F_4}{\partial \vec{\mathbbm{E}}^2}-\frac{\partial F_3}{\partial(\vec{\mathbbm{E}} \cdot \vec{v})}+\frac{\partial F_{10}}{\partial \vec{\mathbbm{E}}^2}\right)\\
    \gamma_{\mathbbm{P},4}&=&\frac{\partial F_6}{\partial(\vec{\mathbbm{E}} \cdot \vec{v})}-\frac{\partial F_8}{\partial \vec{v}^2}-\frac{\partial F_{11}}{\partial \vec{v}^2}+\frac{1}{2}\frac{\partial F_9}{\partial(\vec{\mathbbm{E}} \cdot \vec{v})}\\
    \gamma_{\mathbbm{P},5}&=&\frac{\partial F_6}{\partial \vec{\mathbbm{E}}^2}-\frac{\partial F_7}{\partial \vec{v}^2}+\frac{\partial F_9}{\partial \vec{\mathbbm{E}}^2}\\
    \gamma_{\mathbbm{P},6}&=&\frac{\partial F_7}{\partial(\vec{\mathbbm{E}} \cdot \vec{v})}-\frac{\partial F_8}{\partial \vec{\mathbbm{E}}^2}-\frac{\partial F_{11}}{\partial \vec{\mathbbm{E}}^2}\\
    \gamma_{\mathbbm{P},7}&=&\frac{\partial F_{12}}{\partial(\vec{\mathbbm{E}} \cdot \vec{v})}-F_8-F_{11}\\
    \gamma_{\mathbbm{P},8}&=&2\frac{\partial F_{12}}{\partial \vec{\mathbbm{E}}^2}-2F_7\\
    \gamma_{\mathbbm{P},9}&=&F_6+\frac{1}{2}F_9-\frac{\partial F_{12}}{\partial \vec{v}^2}\\
    \gamma_{\mathbbm{P},10}&=&F_8+F_{11}-\frac{\partial F_{12}}{\partial(\vec{\mathbbm{E}} \cdot \vec{v})}\\
    \gamma_{\mathbbm{P},11}&=&\frac{\partial F_4}{\partial\mu}+\frac{\partial F_{10}}{\partial\mu}\\
    \gamma_{\mathbbm{P},12}&=&\frac{\partial F_8}{\partial\mu}+\frac{\partial F_{11}}{\partial\mu} \; . 
\end{eqnarray}
}

\end{document}